\def\spitzer{\:{\it Spitzer}}
\def\hubble{\:{\it Hubble}}
\def\chandra{\:{\it Chandra}}

\documentclass[12pt,preprint]{aastex}

\begin{document}


\newcommand{\MSOL}{\mbox{$\:M_{\sun}$}}


\title{{\it Spitzer Space Telescope} 
Observations of Kepler's Supernova Remnant: A Detailed Look
at the Circumstellar Dust Component\altaffilmark{1}}
\author{William P. Blair\altaffilmark{2}, 
Parviz Ghavamian\altaffilmark{2},
Knox S. Long\altaffilmark{3},
Brian J. Williams\altaffilmark{4}, \\
Kazimierz J. Borkowski\altaffilmark{4},
Stephen P. Reynolds\altaffilmark{4}
\& Ravi Sankrit\altaffilmark{5}
}

\altaffiltext{1}{Based on observations made with the NASA 
$\spitzer$ Space Telescope. $\spitzer$ is operated for NASA 
by the Jet Propulsion Lab.}

\altaffiltext{2}{Department of Physics \& Astronomy, Johns Hopkins 
University, 3400 North Charles Street, Baltimore, MD 21218}

\altaffiltext{3}{Space Telescope Science Institute, 3700
San Martin Drive, Baltimore, MD 21218}

\altaffiltext{4}{Department of Physics, North Carolina State
University, Raleigh, NC 27695-8202}

\altaffiltext{4}{Space Sciences Laboratory, University of 
California Berkeley, Berkeley, CA 94720-7450}

\begin{abstract}

We present 3.6 - 160 $\mu$m infrared images of Kepler's supernova remnant
(SN1604) obtained with the IRAC and MIPS instruments on the {\it Spitzer Space
Telescope}. We also present MIPS SED low resolution spectra 
in the 55 - 95 $\mu$m region.
The observed emission in the MIPS 24 $\mu$m band shows the entire shell. 
Emission in the MIPS 70 $\mu$m and IRAC 8 $\mu$m bands is seen only 
from the brightest regions of 24 $\mu$m emission, which also correspond
to the regions seen in optical H$\alpha$ images.
Shorter wavelength IRAC images are increasingly dominated by stars, although 
faint filaments are discernible.  The SED spectrum of 
shows a faint continuum dropping off to longer wavelengths
and confirms that strong line emission does not dominate
the mid-IR spectral region.  The emission we see is 
due primarily to warm dust emission from dust heated 
by the primary blast wave;  no excess infrared emission is observed 
in regions where supernova ejecta are seen in X-rays.
We use models of the dust to interpret the observed 70/24 $\mu$m 
ratio and constrain the allowed range of 
temperatures and densities.  We estimate the current mass of dust in the
warm dust component to be $5.4 \times 10^{-4}\ M_\odot$, and infer 
an original mass of about $3~ \times ~ 10^{-3} \ M_\odot$ before grain 
sputtering.  The MIPS 160 $\mu$m band shows no emission 
belonging to the remnant.  We place a conservative but temperature dependent 
upper limit on any cold dust component roughly a factor of 10 below the 
cold dust mass inferred from SCUBA observations.
Finally, we comment on issues relevant to the possible precursor 
star and the supernova type.

\end{abstract}


\section{Introduction}

Each of the remnants of historical galactic supernovae (SNe) provides
a unique and important perspective to our understanding of young
supernova remnants (SNRs) and their interaction with the
interstellar medium (ISM). 
SN~1604 is the second youngest galactic SNR (behind Cas A),
and was first sighted in October 1604 by Johannes Kepler and 
others.  Kepler was not the first to see the SN, but he published the 
most detailed account of the SN light curve (Kepler 1606), and largely 
because of this the SNR has 
come to be known as Kepler's SNR.  Baade (1943) was the first to 
recover the optical SNR. It has been well observed
at all wavelengths (X-ray: Cassam-Chena\"i et al. 2004, Hughes 1999;
Radio: DeLaney et al. 2002; Optical: Blair
et al. 1991, Sollerman et al. 2003; NIR/IR: Gerardy \& Fesen 2001, 
Douvion et al. 2001; sub-mm: Morgan et al. 2003; and numerous earlier 
papers referenced therein).

The distance to Kepler's SNR has been uncertain, with most investigators
adopting a value of near 5 kpc, as discussed by
Reynoso \& Goss (1999).  However, Kepler's SNR lies almost directly
toward the galactic center, making the use of H~I velocities and a galactic
rotation curve very unreliable.  Using an HST/ACS image compared to
ground-based data, Sankrit et al. (2005) have recently determined a
distance of 3.9 (+1.9, -1.4) kpc to Kepler using the proper motion of
a filament whose velocity is known with reasonable accuracy
(Blair et al. 1991).  Since the 
uncertainty is still rather large, we will adopt a value of 4 kpc in 
this paper and scale other parameters to this value. 
The SNR lies $6.8^{\circ}$ off the galactic plane (473 $\rm d_4$ pc).  
With an angular diameter of
$\sim$200\arcsec, the radius is 1.93 $\rm d_4$ pc and the mean expansion 
velocity has been $\rm \sim4720 ~ d_4 ~ km~s^{-1}$.  Since the current 
velocity of the primary blast wave is $\sim$1660 $\rm \pm 120 ~km~s^{-1}$
(Blair et al. 1991; Sankrit et al. 2005), the primary
shock has been significantly decelerated over 400 years.

The brightest optical emission is from a region in the WNW, as shown 
in the images of Blair et al. (1991).  Bright [S~II] emission indicates 
slower, denser, radiative shocks.  However, more extensive nonradiative 
emission is present where smoother H$\alpha$ filaments are not accompanied 
by [S~II], for instance, across the N rim and in portions of two 
centrally-projected regions.  We shall refer frequently below to the 
different character of the radiative and nonradiative emission regions.

Douvion et al.~(2001) observed Kepler with the ISOCAM instrument
aboard the Infrared Space Observatory (ISO), with an angular
resolution of 6\arcsec, in a band from 10.7 -- 12.0
$\mu$m.  Spectra were also obtained between 6.5 and 16 $\mu$m.  They
found the emission to have a similar morphology to the H$\alpha$
image, but made no distinction between radiative and non-radiative
regions.  They found almost no line emission to be present and were able to
fit their spectra with collisionally heated dust at a fixed
temperature, using ``astronomical silicate" composition (Draine \& Lee
1984), assuming that dust was heated in regions with densities and
temperatures consistent with radiative shock emission ($n_e \sim (2 -
10) \times 10^3$ cm$^{-3}$, $T_e \sim 10^4$ K.  They reported a total
dust mass of order $10^{-4}$ $M_\odot$.

Kepler's SNR is the only of the historical SNe from the last millennium
whose progenitor type is a matter of serious debate (Blair 2005).  
The early claim
by Baade (1943) of consistency of the historical light curve with a 
Type Ia SN has been questioned by Doggett 
\& Branch (1985) and Schaefer (1996) among others.  The presence of 
dense, (apparently) N-rich circumstellar material (CSM) surrounding the SNR
so high off the galactic plane was taken to indicate pre-SN mass loss from a
massive star and hence a core collapse SN (Bandiera 1987).
The progenitor could have been a runaway star, which is 
consistent with the observed morphology (Borkowski et al.
1992; 1994; Vel\'{a}zquez et al. 2006).  On the other hand, analyses of 
X-ray spectra from Exosat (Decourchelle 
\& Ballet 1994), ASCA (Kinugasa \& Tsunemi 1999, 2000) and more recently 
XMM-Newton (Cassam-Chena\"i et al. 2004) indicate an overabundance of Fe 
and Si enhanced ejecta, which is not expected from the explosion of 
a massive star but instead suggests a Type Ia designation.

At some point in the evolution of a young SNR, some of the SN ejecta 
may form into interstellar dust of various types, depending on the 
SN type (e.g.  Arendt et al. 1999). Evidence exists for the formation 
of dust on 1-2 year time scales in core-collapse SN (SN 1987A--McCray 
1993 and Bouchet et al. 2004; SN 1999em--Elmhamdi et al.~2003).  
The large mass of iron 
produced by Type Ia SNe suggests that grains may form there as well, 
although direct evidence
for this process is scant and timescales are not well known. 
Dunne et al. (2003) have reported the detection
of a large mass of very cold dust in Cas A based on SCUBA (sub-mm)
observations, although these conclusions have been called into 
question (Dwek 2004a,b; Krause et al. 2004).  Morgan et al. (2003) 
have reported SCUBA observations of Kepler's SNR and claimed a 
large mass of cold (17 K) dust.
A SN blast wave sweeps up gas and dust as it expands, and
so any circumstellar or interstellar dust should also be heated,
and ultimately sputtered and destroyed by collisions in the post shock
region.  Hence, young SNRs can in principle be dust-processing 
laboratories.  With the advent of the {\it Spitzer Space Telescope}, 
it is possible to study these effects in unprecedented detail.

$\spitzer$ observations of Magellanic Cloud SNRs 
are shedding some light on these issues.  Borkowski et al. (2006)
looked exclusively at four remnants of Type Ia SNe.  They found
no evidence of dust from the ejecta, but rather emission consistent
with heating of ISM dust by the expanding blast wave.  B. Williams
et al. (2006) selected a group of suspected remnants of
core collaspe SNe in the LMC and have found similar results: no
obvious dust emission associated with regions of ejecta but
dust emission associated with the primary blast wave was
prominent.  R. Williams et al. (2006) report $\spitzer$
observations of a different sample
of LMC SNRs and suggest that, at least in some objects,  
line emission may dominate over the dust continuum.  However,
even in these objects, the emission arises at positions consistent
with shocked ISM.

In this paper, we report infrared imaging and spectroscopy of 
Kepler's SNR using 
the MIPS and IRAC instruments on $\spitzer$, obtained as part of a
Cycle 1 Guest Observer program (\#3413). We find evidence for swept-up
dust heated by the primary blast wave, but do not see evidence for
the cold dust component reported by Morgan et al. (2003).
Modeling of the dust and comparison to observed ratios are 
used to constrain the allowed plasma conditions in the shocked gas.

\section{Observations and Data Processing}

Below we describe the $\spitzer$ MIPS (Rieke et al. 2004) and IRAC 
(Fazio et al. 2004) imaging and present the data.
The MIPS instrument also has a low resolution spectral energy distribution
(SED) spectroscopy mode, and the data obtained on Kepler are described 
in the subsequent sub-section.

\subsection{$\spitzer$ Imaging}

We observed Kepler's SNR using MIPS imaging at 24, 70 and
160 $\mu$m, and with all four bands of IRAC (3.6, 4.5, 5.8 and 8.0
$\mu$m).  All MIPS photometric imaging used the `small' field size. 
The 160 $\mu$m band should be most sensitive to the distribution
of cold dust, being near the peak of the curve fit by Morgan et al.
(2003) to the SCUBA data.  Based on the ISO measurements of Douvion et al.
(2001), the 24 and 70 $\mu$m MIPS bands and 8 $\mu$m IRAC band should
assess the warmer dust component near 120 K.

At 24 $\mu$m we used one cycle and no mapping with a 10 s exposure 
time.  The 70 $\mu$m band used 10 s exposure times, 1 cycle
and a 3x1 map for two cycles (step size at 1/2 array for columns
and full array for rows).  This results in 252 sec integration time
and covers the object with room to spare to assess background,
and enough overlap to account for the side B problem.
The smaller FOV of the 160 $\mu$m band (along with the
anomalous block of 5-pixels) required a greater number of positions
to cover the desired region. We chose 10 sec exposures and 4 cycles.
We then chose mapping with 1x3 for 3 map cycles.  This results in
an integration time of 252 sec. 

We worked with the standard post-BCD reduction data sets for 
the MIPS imaging data, as retrieved with {\tt Leopard} from the 
$\spitzer$ archive.  The data have been reprocessed several times over
the course of this work, and here we report the S11.0.2
version of the processed data.  Figure 1 shows the full field of 
view MIPS 24 $\mu$m image, which is by far the most sensitive and
detailed of the $\spitzer$ images obtained.

Kepler fits within a single IRAC field. Our AOR used the 12
position Reuleaux dither pattern and 2-100 sec frames per pointing to
to achieve a 2400 sec integration time in all four bands. 
We again worked with the post-BCD data sets, which were processed with
version S11.4.0.  While some SNR emission is apparent in the 8 $\mu$m 
image, it is difficult to assess the shorter wavelength IRAC images
for SNR emission because of stellar contamination. 

Hence, to better investigate the extent of SNR emission in the IRAC bands,
we have displayed these data in two ways.  First, we made three-color
images.  Figure 2a shows a three-color
IRAC image with 8 $\mu$m (red), 5.6 $\mu$m (green), and 3.6 $\mu$m (blue).
Stars are white or bluish and the SNR emissions appear orange and yellow.
Orange filaments show emission at both 8.0 and 5.6 $\mu$m, but are dominated
by the longer wavelength band.  Yellow filaments appear to correspond
closely with the brighter radiative filaments and indicate relatively 
stronger emission at 5.6 $\mu$m. Any SNR emission at 3.6 $\mu$m is 
completely dominated by the longer wavelengths. Figure 2b shows a 
similar comparison, but using 5.6 $\mu$m as red, 4.5 $\mu$m in green 
and again 3.6 $\mu$m in blue. Even more so than Fig. 2a, this combination 
shows only emission from the densest (radiative) optical filaments, which
appear as orange (indicating emission in both the 4.5 and 5.6 $\mu$m bands).
The two patches of green emission in this Figure adjacent to bright stars
indicate instrumental effects in the 4.5 $\mu$m image.

Secondly, we have made difference images of adjacent bands to minimize
the effects of stellar contamination.  Since SNR emission likely occurs 
in the band being used for subtraction, these images are not useful for 
quantitative analysis, but rather for highlighting the overall extent of
detected emission. 
Fig. 2c shows the residual emission when the 5.6 $\mu$m band is subtracted 
from the 8 $\mu$m image, and Figure 2d shows 4.5 $\mu$m minus 3.6 $\mu$m.
While some stellar residuals remain, these two panels confirm the discussion 
from the three-color Figures above: fairly extensive emission is present 
at 8$\mu$m, and only a few of the brightest clumpy filaments remain visible
at 4.5 $\mu$m.

For comparison, we show a medium scaling of the MIPS 24 $\mu$m image in
Figure 2e, and a 0.3 - 0.6 keV energy cut of archival {\it Chandra} soft
X-ray data in Figure 2f. Comparing Figure 2c and 2e shows that the region
visible at 8 $\mu$m corresponds very closely to the brightest regions
in the 24 $\mu$m image. There is also tremendous similarity between the
softest X-rays and the 8 $\mu$m image. Since the densest regions should be
coolest, this confirms to first order that the brightest regions in the
8 and 24 $\mu$m images are largely due to higher densities at these locations.

In Figure 3, we show the MIPS data and some additional comparisons, all 
on the same scale as the panels of Figure 2. Figure 3a and 3b show hard 
and soft stretches of the 24 $\mu$m data, respectively, highlighting the 
faintest and brightest regions detected at 24 $\mu$m.
The MIPS 70 $\mu$m data were reprocessed
using the GeRT software available from the $\spitzer$ Science Center
web site in an attempt to minimize the obvious striping that
traverses these data from NNE to SSW.  This made a modest improvement
in the cosmetic appearance of the 70 $\mu$m image, but the original
data set was used for all measurements reported in this paper.
The corrected data are shown in Figure 3c. Although the resolution is
lower at 70 $\mu$m, it is quite clear that the regions brightest at
70 $\mu$m correspond closely to the brightest emission at 24 $\mu$m.
Figure 3d shows the star-subtracted H$\alpha$ image from Blair et al. 
(1991). The overall extent of the H$\alpha$ emission is quite silimar
to the brightest infrared regions, although differences are apparent
in the brightest region of radiative filaments in the WNW.  Figure 3e
shows the MIPS 160 $\mu$m image of the same region as the other panels.
No emission is detected above the complex and variable background at 
160 $\mu$m.  Finally, Figure 3f shows a three color rendition of the
{\it Chandra} X-ray data. The red shows the same energy cut as in Figure 2f
(0.3 - 0.6 keV), green is the energy band 0.75 - 1.2 keV, and blue is
a band from 1.64 - 2.02 keV.

\subsection{MIPS SED Spectroscopy}

The MIPS SED spectroscopy mode covers the spectral region from 55 $\mu$m
to 95 $\mu$m with a resolution of 15 - 25.  The SED aperture covers a 
region 3.8\arcmin\ $\times$ 0.32\arcmin\ and produces a two-dimensional 
spectral
output file with 16 pixels in the spatial direction and 32 pixels in
the dispersion direction.  A grid of aperture positions covering
Kepler's SNR and an adjacent sky region were observed on 25 Sep. 2005.
The spectral grid used a 3\arcmin\ chop distance in the dispersion
direction and an overlapping ``a" and ``b" position in the spatial
direction, covering the northern 2/3rds of the shell (mapping with two
overlapping columns
and 7 rows).  With five cycles, this provided $\sim$300 sec integration times
per position.

The primary purpose of the SED spectra was to search for evidence that
strong emission lines might be contaminating the broad band imaging,
and thus affect their intercomparison. The individual spectra are
low in signal and so we chose to sum the data covering the bright NW
quadrant of the remnant (grid positions 2b, 3b, 4b, and 5b), using 
the corresponding sky positions that were furthest removed from SNR  
emission (sky positions s4b, s5b, s6b and s7b). The aperture positions
used are shown in Figure 4, projected onto the 24 $\mu$m image.
After summing the object data and subtracting sky, it was clear from
displaying the two-dimensional data that the strongest signal was
in columns 8 - 12.  We collapsed these four columns into the
spectrum shown in Fig. 5, which represents
the bright NW radiative filaments and a portion of the NW shell.
Because of the relatively crude state of SED calibration and
software for handling these data, we do not attempt quantitative
fits to the resulting spectrum, but rather rely on it to provide
a more general description of the mid-IR spectrum.

The SED spectrum is dominated by the tail of the warm dust 
continuum as it fades to longer wavelengths.  There is no evidence at
this spectral resolution for the [O~I] 63.2 $\mu$m line predicted to
be strong in radiative shock models.  There may be some indication of
the 88.3 $\mu$m line of [O~III], but it does not dominate the spectrum.
A similar bump near 75 $\mu$m may be due to [N~II] 76.5 $\mu$m,
which might be consistent with the N overabundance of the shell 
material as judged from optical spectra, although the wavelength 
agreement is not very good. In any event, since lines do not dominate
in this spectral region where the dust continuum is relatively faint,
it is unlikely that line contamination has a signficant effect on the
observed 24 $\mu$m flux where the dust emission is much stronger.

This point is strengthened by an IRS spectrum of the
NW region by Roellig \& Onaka (2004).  This spectrum shows 
a modest emission feature near 26 $\mu$m that they mark as [Fe II]
but which may be a blend of this line with [O IV] 25.9 $\mu$m. 
However, this emission feature makes only a small contribution to the
total flux in the 24 $\mu$m band.  The IRS spectrum extends down
to the IRAC range, showing the dust continuum almost disappearing.
In the IRAC 8 $\mu$m bandpass, a moderately strong [Ar II] line
appears at 7.0 $\mu$m, accounting for $\sim$ 20\% of the total
flux.  Hence, this is potentially  important in considering ratios 
between the 8 $\mu$m image and other bands, at least for the NW filaments 
where bright radiative emission is dominating (e.g. Blair et al. 1991).

\section{Analysis and Modeling}

\subsection{Morphological Comparisons}

What is the structure of Kepler's SNR in the infrared and how does
it compare to images at other wavelengths?  
Although the effective resolutions in the various data sets are different,
they are close enough to allow some meaningful intercomparisons.
The 24 $\mu$m data are the deepest of the $\spitzer$ data. Scaling to show
only the highest surface brightness features, the similarity of Fig. 3b
to the IRAC 8 $\mu$m (Fig. 2c) and 70 $\mu$m (Fig. 3c) indicates to 
first order that the same regions dominate all three bands.
The further similarity to the optical
H$\alpha$ image (Fig. 3d) and the softest band of X-ray emission 
(Fig. 2f and red band in Fig. 3f) is striking.
Since the majority of the optical 
emission especially across the northern limb  arises from nonradiative 
shocks associated with the primary blast wave, it is clear that the 
brightest 8 $\mu$m, 24 $\mu$m, and 70 $\mu$m  emissions are associated 
with this same component.  Hence, heating of dust in dense, swept up CSM/ISM 
is the dominant mechanism operating in Kepler's SNR. 

The morphology of the WNW region including the primary region of
bright radiative filaments contains some subtle but important effects.
Careful comparison of Fig. 3b and Fig. 3d shows differences between 
the brightest 24 $\mu$m morphology and the chaotic, more extended
emission seen in the H$\alpha$ image. We suggest the 24 $\mu$m image is
dominated by emission from the primary blast wave in the moderate density 
interclump gas, and not the slower
radiative shocks in the very dense radiative knots, which are dominated
by [S~II], [N~II], and [O~III] line emission (e.g. Blair et al. 1991).  
The 8 $\mu$m image (Fig. 2c)
looks intermediate between the 24 $\mu$m and H$\alpha$ appearance,
which is another indicator that line emission from the radiative shocks
contributes somewhat at this wavelength, as indicated above in the discussion of
the IRS spectrum.  

The 24 $\mu$m image scaled to highlight the lowest surface brightness 
structures (Fig. 3a) is the only of the IR images with sufficient signal 
to show the entire outer shell of the SNR.  A general similarity is seen with
the 3-color X-ray image in Fig. 3f, with the exception of the extended
`ears' to the WNW and ESE.  Similarities to the 6 cm VLA data are apparent 
(e.g.  DeLaney et al. 2002), in 
particular in the south and along the eastern rim.  However differences
are also apparent, including a relatively higher IR surface brightness
in the NW, along the northern rim, and across the projected middle.
The `ears' mentioned above are most apparent in the radio data.

The disagreement in appearance in the projected interior of the shell
deserves specific comment.  Here an important clue comes from the optical
spectroscopy of these central filaments by Blair et al. (1991).
The SE central grouping of filaments are significantly red-shifted and thus
are a portion of the receding shell of the SNR.  In comparison, the
NW central filaments are blue-shifted and are part of the approaching
shell of the SNR. These filaments have direct counterparts in the soft
X-ray band, but the association with the higher energy X-ray bands is
less clear. The approaching filaments have no counterpart in
the radio data, and while there is faint radio emission at the projected
position of the SE central grouping of filaments, the morphologies 
are different.  Hence, it is not clear whether any of the central radio 
emission is correlated with either the optical or bright IR emissions 
in these regions.

In Figure 6, we show a different kind of morphological comparsion. This
color composite, star-subtracted image shows Kepler's SNR as viewed by NASA's
three Great Observatories, with $\spitzer$ 24 $\mu$m data in red, $\hubble$
H$\alpha$ in yellow, $\chandra$ 0.3 - 1.0 keV emission in green, and
harder 2 - 10 keV $\chandra$ emission in blue.  This image provides a
combination of physical information and some subtle affects due to differing
spatial resolution of the data sets used.  The $\hubble$ data have the
highest spatial resolution, but highlight the fact that the optical emission
arises in knotty, dense structures being encountered by the primary blast 
wave.  These filaments are bathed in the glow of the IR component, which 
arises from dust heated by the blast wave.  The X-ray emission in the 
north extends out to the optical limb and stops.  The red rim 
around the top is largely an artifact of the lower resolution of the 24 
$\mu$m data and the stretch applied to this component of the image. (As 
we indicated in the discussion above, the brightest 24 $\mu$m emission 
is coincident with the shock front position.)
However, the larger extension of the red emission in the NW appears to 
be a real effect.

The distribution of the two X-ray components in Fig. 6 is quite 
interesting.  The harder X-rays (blue) primarily arise in sychrotron 
emission at the
shock front, and are seen most clearly along the southern and eastern limbs.
The softer X-ray component is dominated by thermal emission from Si and Fe
rich ejecta (e.g. Cassam-Chena\"i et al. 2004). In the south, this component
lies directly interior to the blue component.  In the north, this component
largely fills in between the northern limb emission and the 
centrally-projected emission regions. Note the distinct absence of red 
emission in the green regions, which indicates {\it no significant 
warm dust emission} coincident with regions of ejecta.

\subsection{Derivation of IR Fluxes and Ratios}

\subsubsection{Total Fluxes at 24 $\mu$m and 70 $\mu$m}

We now turn to more quantitative information, concentrating first on
the 24 $\mu$m and 70 $\mu$m data sets, where contamination of the SNR 
emission by stars is not a significant problem.  We derived the total fluxes
at these two wavelengths using the following method.  Because the post-BCD
data are in units of MJy $\rm sr^{-1}$, we extract regions of pixels 
corresponding to the object, and use 
the known post BCD pixel sizes (from the file headers) and number of 
pixels included in each region to scale appropriately to total fluxes.  
We similarly extract representative 
regions of background surrounding the SNR region (as allowed by field 
coverage) and average these to determine the most appropriate overall 
background levels to subtract.  Using this technique, we obtain total
fluxes at 70 and 24 $\mu$m of 4.90 Jy and 9.5 Jy, respectively, and
thus a ratio of 70/24 $\mu$m of 0.52.  

Our total flux at 70 $\mu$m appears to disagree with published 
{\it IRAS} fluxes at
60 $\mu$m (range from 7.1 - 10.5 Jy), summarized by Saken et al. 
(1992, their Table 6), although our derived 24 $\mu$m flux lies
within the range listed (8.1 - 11.7 Jy) at IRAS 25 $\mu$m. 
The apparent disagreement arises simply from the
differing bandpasses used.  Bandpasses for {\it IRAS} are broader 
than for {\it Spitzer}, and thus would encompass more flux at a 
given wavelength.  To confirm this, we integrated the output 
spectrum from the dust model described below for the entire remnant 
over the {\it IRAS} 60 $\mu$m bandpass and obtained an ``expected" 
60 $\mu$m flux of $\sim$8 Jy.  A corresponding exercise for the
{\it Spitzer} 70 $\mu$m band predicts a flux of ~5 Jy. 

\subsubsection{Flux Measurements for Regions}

In addition to the total fluxes, we have determined the 24 and 70
$\mu$m fluxes for several sub-regions of the remnant to search for
variations in these ratios that might arise due to different shock or
other parameters.  We have also measured fluxes from relevant regions
of the IRAC 8 $\mu$m image, although stellar contamination is a
more significant problem for these data.  Figure 7 shows 
the extraction regions projected onto the relevant images.
These figures also define the nomenclature we will use below
to reference the regions.

Because {\it Spitzer}'s optics provide images at or near the
diffraction limit for all wavelengths, 
the resolution in images from various instruments differs by roughly the
ratio of wavelengths.  The angular resolutions of
various images range from  2\arcsec\ at 8 $\mu$m, to 6.2\arcsec\ at 
24 $\mu$m, 18\arcsec\ at 70 $\mu$m, and 41\arcsec\ at
160 $\mu$m.  For determination of flux ratios from various
regions of the SNR, it is
necessary to convolve the higher-resolution image of interest to the 
resolution of the lower-resolution image.  For the 8/24 $\mu$m ratios,
we convolve the 8 $\mu$m image to the resolution of that at 24 $\mu$m,
and for the 24/70 $\mu$m ratios, we convolve the 24 $\mu$m image to 
the resolution of that at 70 $\mu$m.  We used contributed software by
K.~Gordon (U.~Arizona) distributed by the SSC to convolve the
images, using  kernels for convolving
24 $\mu$m images to the same PSF as 70 $\mu$m, and similarly for 8
$\mu$m to 24 $\mu$m.  These kernels are slightly temperature-dependent;
we used versions appropriate for 100 K.  Though we do not expect
blackbody spectral shapes, this value is close to the appproximate
grain temperatures reported by Douvion et al. (2001).  In each case, 
the higher-resolution images were
resampled onto a grid identical to that of the lower resolution
image, using AIPS
(Astronomical Image Processing System)\footnote{ AIPS is produced and
supported by the National Radio Astronomy Observatory, operated by
Associated Universities, Inc., under contract with the National
Science Foundation.}  task HGEOM.  These are the images shown in
Figure 7.

The extraction regions were defined to select physically associated
regions of emission so that the results could be compared.
We displayed and aligned the 24 and 70 $\mu$m images using the display
tool {\tt ds9}. \footnote{See {\tt http://hea-www.harvard.edu/RD/ds9/} .}  
We also displayed the optical H$\alpha$
and other images on the same scale for reference and comparison 
while defining regions. 
The defined 24 and 70 $\mu$m regions are shown
in Figure 7a-c, and details are given at the top of Table 1. 
Object position O1 enclosed the
region of bright radiative emission in the NW, position O2 enclosed
the entire bright NW rim at 24 $\mu$m. Note that by differencing 
these two positions, the area corresponding to just the relatively bright
nonradiative portion of the NW rim can be extracted.  O3 samples the 
northern (mainly) nonradiative rim, and O4 and O5 enclose the two 
centrally-projected regions of knotty filaments.  

Background regions were identified to the north and south of the remnant to
account for the gradient observed in the background from north to south.
Some backgrounds were defined specifically for use with the 70 $\mu$m 
image in an attempt to better account for the significant striping in
this image. For each object position, the most appropriate 
backgrounds were summed and scaled to the size of the object region 
being measured.  Specifically, we used the following regions as background 
at 70 $\mu$m: For positions O1 and 
O2 in Figure 7a-c, background regions B1 and B5 were averaged and used. For 
positions O3 and O5, background regions B2 and B4 were used. And for position 
O4, background regions B2 and B3 were used.  At 24 $\mu$m we simply used
regions B1, B2 and B4, which do not overlap SNR emission.

The signal in each sampled region was measured using the {\tt FUNtools}
package that interfaces with 
{\tt ds9}. \footnote{{\tt http://hea-www.harvard.edu/RD/funtools/help.html} .}
As with the total flux estimates, it is necessary to multiply the total 
signal above background measured in a region by the pixel scale 
(sr $\rm pixel^{-1}$). 
Thus, ratios use the same spatial object regions, even though the pixel 
sizes varied. The region fluxes and ratios between 24 and 70 $\mu$m are
provided in Table 2.  Only small differences are obtained.  In particular,
the similarity in ratio between O1 and O3, or O1 and the difference between
O2-O1, indicates that there is no significant difference in 70/24 $\mu$m
ratio between bright radiative and nonradiative regions.  This is in 
keeping with the earlier discussion about the IRS and SED spectra, where
these emissions are dominated by the main blast wave.

We have also derived ratios of regions between 8 $\mu$m and 24 $\mu$m
using a similar technique, although for somewhat modified spatial 
regions than used with the 70/24 $\mu$m ratios.  
The defined 8 and 24 $\mu$m regions are shown
in Figure 7d-f, and details are given at the bottom of Table 1. 
This comparison is complicated by the faintness of the 8 $\mu$m emission, 
by the presence of many more stars at 8 $\mu$m, and by the difference in 
resolution between the images.  
Because of potential contamination of the 8 $\mu$m image by the 
[Ar~II] 7.0 $\mu$m line seen by Roellig \& Onaka (2004), the 
derived ratios may be skewed toward slightly higher 8/24 $\mu$m
ratios than true.  On the other hand, any relative changes in the ratio
should be real. 


For the 8 $\mu$m to 24 $\mu$m comparison,
positions O1 and O2 isolate the two brightest 
regions of radiative optical filaments in the NW.  
Position O3 samples the so-called `bump' region, which includes
both radiative and nonradiative optical emission in the NNW.  O4 provides 
the cleanest sampling of nonradiative shock emission on the northern rim.
O5 and O6 sample two regions of central emission knots, but are smaller 
regions than measured above for the 70/24 $\mu$m ratio. Again, optical 
data suggests a mixture of radiative and nonradiative shocks in these regions.
Stars have largely been avoided, with the exception of position O6, 
for which it was impossible to totally exclude stars from the selected 
region.  At 8 and 24 $\mu$m, we simply averaged all three of the 
background regions shown and applied this as representative
to all positions.

We have applied recommended photometric corrections for diffuse
sources appropriate for the 8 $\mu$m band, as described in the IRAC
section of the SSC website\footnote{See {\tt http://scc.spitzer.caltech.edu/irac/} .}.  
This correction for sources large compared to the
calibration aperture for point sources (12\arcsec) is a factor of 0.74,
which has been applied to measured fluxes and backgrounds for this band. 
However, this correction is sufficiently poorly known that an
additional uncertainty of order 10\% is introduced. Any relative 
changes in the 8/24 $\mu$m ratio should remain valid, however.

The derived 8/24 $\mu$m ratios are summarized in Table 3.
Comparing the measurements at positions O1 and O2 to O4, we see
higher ratios at the strongly radiative positions, indicative of 
contamination of the 8 $\mu$m image by line emission (likely [Ar II], 
as indicated in the IRS spectrum discussion above).  Hence, the O4 
measurement likely represents the most accurate assessment of the 
8/24 $\mu$m dust continuum. It is unclear without IRS data whether 
the higher ratios observed at O3, O5, and O6 are due to variations 
in the dust continuum or due to significant IR line emission from the 
optically faint radiative filaments in these regions. 

We have not color-corrected the observed fluxes reported in the
Tables.  Our model fluxes have been produced by integration of the
calculated spectrum over the {\it Spitzer} bandpasses at 24 and 70
$\mu$m.  Since {\it Spitzer} fluxes are calibrated by comparison with
stars whose spectra at 24 $\mu$m and longward are well-approximated by
the Rayleigh-Jeans limit of a blackbody, we have assumed a
$\lambda^{-2}$ spectrum and calculated what {\it Spitzer} would report
for the flux at the nominal frequency for each band.  That is, we have
converted our model fluxes into ``{\it Spitzer} space'' before
computing model ratios, rather than color-correcting observed fluxes.
However, such corrections would not be large in any case.  Our shock
models that reproduce the observed flux ratios give grain temperatures
between 75 and 95 K.  While the spectra are not exact blackbodies,
color corrections for a 70 K blackbody are less than 10\% at both 24
and 70 $\mu$m, as reported in the MIPS Data Handbook (v3.0, p.~29); at
100 K, they are less than 7\% at both wavelengths.  For IRAC, color
corrections are reported in the IRAC Data Handbook only down to
blackbody temperatures of 200 K.  For that temperature, corrections
are less than 20\% for both Channels 3 and 4.  We believe that overall
calibration errors, estimated in the {\it Spitzer} Observing Manual,
Sec. 8.3.3 (p. 355) to be 10\% for extended sources at 24 $\mu$m and
15--20\% at 70$\mu$m, will dominate the errors at those bands.  We can
also estimate internal statistical errors from the pixel-to-pixel
dispersion in the background; these errors are negligibly small
compared to those due to calibration.  We adopt conservative overall
error estimates of 10\% at 24 $\mu$m and 20\% at 70 $\mu$m. 

In addition to deriving region ratios as
described above, we also created a ratio map from the convolved
24 $\mu$m image and the 70 $\mu$m image as follows.  First, we
measured a mean background from most of the region on the image not
occupied by the source (avoiding obvious point sources), and
subtracted that value, at each wavelength.  Then we blanked the
convolved 24 $\mu$m image below 10 MJy/sr, a level corresponding to
about 20\% of peak, which left all obvious structure intact.  (The
off-source rms fluctuation level was 0.6 MJy/sr.)  At 70 $\mu$m, a
much noisier image, the off-source rms was about 1.6 MJy/sr; we
blanked below 5 MJy/sr, about three times this value.  The point of
the blanking is to make sure that only pixels whose measured fluxes
are highly significant at both wavelengths are used to compute ratios.
We then generated the ratio image shown in Fig.~8.

Figure 8 shows several features of interest.  First, a general
anticorrelation of ratio with 24 $\mu$m brightness indicates that
hotter (lower 70/24 $\mu$m ratio) regions are brighter.
Second, the minima in the ratio are actually offset from the 24 $\mu$m
brightness peaks slightly.  Third, the western region
of bright radiative shocks appears to have somewhat different ratio.
The range of pixels in the image is about 0.3 -- 0.8, with a
broad maximum around 0.45 and most pixels between 0.35 and 0.5,
consistent with the values measured in regions shown in
Table~2. Higher ratios in the fainter regions is the primary
reason the total 70/24 $\mu$m = 0.52 even though most of the brighter
regions have lower ratios.

\subsection{Grain Emission Modeling}

The morphological comparisons across different wavelengths show a clear
correlation between the soft X-ray images and IR images at all three MIPS
wavelengths, 8, 24
and 70 $\mu$m. We thus attribute emission from Kepler in these bands
to shocked interstellar and circumstellar dust, heated by the hot, X-ray 
emitting plasma in the primary blast wave (Dwek \& Arendt 1992). 
Collisions with energetic electrons and
ions heat dust grains to $\sim 100 K$, where they emit thermal
radiation visible to {\it Spitzer's} mid-IR instruments. In addition
to heating, the ions in the plasma sputter dust grains, rearranging
the grain-size distribution by destroying small grains and sputtering
material off of large grains. We employ computer models of
collisionally heated dust to explain what is seen in Kepler, identical
to what was done for Type Ia SNRs (Borkowski et al. 2006) and
core-collapse SNRs (B. Williams et al. 2006) for SNRs in the 
Large Magellanic Cloud.  Our models use as input an
arbitrary grain-size distribution, grain type (astronomical silicates,
carbonaceous, etc.), proton and electron density $n_p$ and $n_e$, ion
and electron temperature $T_i$ and $T_e$, and shock age (or sputtering
time scale) $\tau_p=\int_0^t n_p dt$. The model is based on the code
described by Borkowski et al. (1994) in the context of photon-heated 
dust in planetary nebula Abell 30, and augmented to allow for heating by 
energetic particles in hot plasmas.

Because little is known about the surroundings of
Kepler, we adopt a power-law grain size distribution with 
index $\alpha$ = -3.5 and an exponential cutoff.  We use a range of
100 grain sizes from 1 nm to 0.5 $\mu$m. We also use only astronomical
silicates.  
This is consistent with past efforts to model dust emission from Kepler
(Douvion et al. 2001).  We use bulk optical constants for astronomical
silicates from Draine \& Lee (1984). Energy deposition rates for
electrons and protons were calculated according to Dwek (1987) and
Dwek \& Smith (1996). Because small grains are stochastically heated
and have large temperature fluctuations as a function of time, we must
account for the increased radiation produced by such transient
heating. We use the method described by Guhathakurta \& Draine (1989)
for this purpose. 

It is also necessary to model sputtering for all grains, since 
sputtering alters the grain size distribution downstream of the
shock.  Sputtering rates for grains in a hot plasma are taken from 
Bianchi et al. (2005). Small grains can actually experience an enhancement
in sputtering due to the ion knocking off atoms not only from the front
of the grain, but also from the sides or the back. We have included 
such enhancements in our models, with enhanced sputtering yields  
described by Jurac et al. (1998). 
Sputtering is very important in the dense CSM environment of Kepler, and
results in the efficient destruction of small grains in the postshock gas and
very significant modification of the preshock grain size distribution. 
Even MIPS fluxes and their ratios depend on the shock sputtering age, but
because emission in these bands is mostly produced by relatively large grains 
with moderate temperature fluctuations, sputtering effects are less extreme 
than at shorter wavelengths. We consider our dust models reliable for modeling 
MIPS fluxes and their ratios. Thermal fluctuations are particularly important 
in very small grains, which reradiate their energy at short wavelengths, so 
sputtering dramatically reduces the amount of radiation in the IRAC bands.
 
Contrary to expectations, the 8 $\mu$m emission is much stronger 
than predicted by our models.  While we do not include PAHs in our 
modeling code, they are not likely to be present in large quantities
given the absence of their distinctive spectral features in IRS
spectra (Roellig \& Onaka 2004).  However, the model does not account
for several physical effects likely to be important for very small grains,
such as discrete heating (i.e., discrete energy losses as an
electron or proton traverses a grain), and corrections to heating and
sputtering rates required when particle mean free paths are much
greater than grain radius.  For these reasons, we have not attempted
to model the 8/24 $\mu$m flux ratios shown in Table~3. A
great deal of theoretical work remains to be done before the wealth of
data produced by {\it Spitzer} can be used effectively to understand
the properties of small grains. 

For modeling an outward moving shock wave, we have used a one-dimensional
plane-shock approximation. The plane-shock model assumes a constant
temperature, but superimposes regions of varying sputtering timescale
from zero up to a specified shock age (Dwek et al. 1996). The shock model
effectively varies the product of density and time behind the shock,
allowing us to account for material that has just been shocked and
material that was shocked long ago, since these will experience different
amounts of sputtering. The output of our models is a single
spectrum, which is produced by superimposing spectra of many different
grain sizes.  Since we
do not model observed spectra directly, we can only fit flux ratios
from photometric measurements. We focus here on reproducing the
observed 70/24 $\mu$m ratios.

We constructed a grid of dust models
to explore the parameter spaces of temperature and density, covering
electron temperatures $kT_e$ from 0.03 to 10 keV and
electron densities $n_e$ ranging from 3 to $10^{3}$
cm$^{-3}$.  We assumed that ion and electron temperatures are
equal. The output of the grid was $\sim 300$ separate models with varying
$n$, $T$, and $\tau_p$, the shock sputtering age. (Since $\tau_p$ inherently 
contains $n_p$, and $n_p$ is related by a constant factor to $n_e$, $\tau_p$ 
is varied from model to model.  See Borkowski et al. 2006.)  We divide the 
actual age of Kepler by 3 
to approximate an ``effective shock age" for Kepler, which 
was used in the models. The factor of $1/3$ arises from applying 
results of a spherical blast wave model to the plane-shock calculation 
(Borkowski et al. 2001). 

For each of the 300 models, a 70/24 $\mu$m ratio was calculated 
from the output spectrum, and the value of the ratio was plotted on a
two-dimensional color-coded plot as a function of electron density and 
pressure (see Figure 9). We then added
contours to the plot which correspond to measured values of the 70/24
$\mu$m ratio from the MIPS images. The three contours are the highest
and lowest measured region values (regions O3 and O5, 0.40 and 0.30, 
respectively; see Table 2), and the value of 0.52 appropriate for 
the spatially-integrated 
fluxes. The plasma conditions for the nonradiative
regions of Kepler's SNR can be contained for the most part between these 
contours. Our nonradiative shock models are not applicable at high densities 
and low temperatures (toward the lower right corner of Fig. 9) because 
of the onset of radiative cooling. Shocks with an age of 400 yr and 
solar abundances. for instance, become radiative to the right of 
the line shown in the lower right corner of Figure 9. 
We used postshock cooling ages tabulated by Hartigan et al. (1987) to draw 
this line in the electron temperature--pressure plane.  

At low densities
(near the left boundary of Fig. 9), our models with equal ion and
electron temperatures overestimate the 70/24 $\mu$m flux ratio. 
In this region of plasma parameters, shocks are fast and ion 
temperatures are generally larger than electron
temperatures. Heating of grains by ions becomes relatively more important, 
resulting in increased grain temperatures and lower 70/24 $\mu$m flux ratios 
relative to shocks with equal ion and electron temperatures.
We estimated an electron temperature for the northwest portion 
of the remnant, where Sankrit et al. (2005) determined 
a shock speed of 1660 $\rm km~s^{-1}$. We used a simple
model for the ion-electron equilibration through Coulomb collisions 
behind the shock. From this we derived a $T_e$ of 1.2 keV (in the 
absence of significant collisionless electron heating and assuming 
a shock age of $\sim 150$ yr).  A shock model
with a postshock electron density of 22 cm$^{-3}$ reproduces the measured 
ratio of 0.40; we mark its position by a star in Figure 9. (Without 
sputtering, this ratio would have been equal to 0.31.) A noticeable 
displacement of this model from the middle contour is caused by additional
heating by ions in the more sophisticated model with unequal ion and 
electron temperatures used for the Balmer-dominated shock in the 
north. Grain temperatures in this model vary from 75 K to 95 K.

Our estimate of a typical electron density in dust-emitting regions
of $n_e \sim 20$ cm$^{-3}$ is in reasonable agreement with other
estimates, such as the estimate of $7 - 12$ cm$^{-3}$ in the
central optical knots (Blair et al.~1991).  It should characterize 
the bulk of the shocked CSM
around Kepler, although higher postshock electron densities and lower
temperatures (and hence lower shock speeds) are also possible because 
contours of constant 70/24 $\mu$m ratio approximately coincide with lines of 
constant pressure on the right hand side of Figure 9.  While nonradiative 
Balmer-dominated shocks with speeds less than 1660 km s$^{-1}$ have 
not been measured to date in Kepler, the highly inhomogeneous 
optical and X-ray morphologies suggest
that plasma conditions in the shocked CSM may vary greatly with 
position within the remnant. It is likely that shocks are present 
with velocities less than seen in the Balmer-dominated shocks but 
more than seen in radiative shocks. Shocks with such
intermediate velocities are best studied at X-ray wavelengths, 
and a future analysis of the CSM in Kepler based on a new long 
{\it Chandra} observation is in progress (Reynolds et al. 2006). 
It is also likely that fast nonradiative shocks in the south and east
travel through gas with densities much less than 20 cm$^{-3}$;
the current MIPS 70 $\mu$m data and optical observations are not 
sensitive enough to study these shocks in much detail.

Our derived value of $n_e T_e = 3 \times 10^{8}$ cm$^{-3}$ K ($n_e
kT_e = 26$ cm$^{-3}$ keV) is in the pressure range considered by Douvion
et al. (2001). However, based on the morphological resemblance between 
the 24 $\mu$m image and the nonradiative optical emission, a
lower density and higher temperature better characterize the typical
dust emitting regions.

\subsection{Total Dust Mass and Dust/Gas Ratio}

The spatially-integrated IR spectrum of Kepler is produced by grains of 
widely varying sizes immersed in inhomogeneous plasmas. This results in a wide 
range of grain temperatures, making determination of a total dust mass 
model dependent because of the extreme sensitivity of the radiated IR power 
to the grain temperature. In Appendix A, we estimate the shocked CSM 
X-ray emission measure in Kepler using a simple plane shock model with 
a mean temperature 
of 5 keV and ionization age of $10^{11}$ cm$^{-3}$ s, without any collisionless 
heating at the shock front but allowing for energy transfer from ions to 
electrons through Coulomb collisions. Emission measure-averaged ion and 
electron temperatures are 8.9 keV and 1.4 keV in this shock model. The only
remaining free parameters in the model are a postshock electron density $n_e$ 
and a total dust mass. We obtain $n_e = 13$ cm$^{-3}$ by matching the measured 
spatially-integrated  70/24 $\mu$m flux ratio of 0.52.
(There is a nonnegligible contribution to grain heating from hot ions in this 
fast shock; as can be inferred from Figure 9, simpler shock models with equal 
ion and electron temperature of 1.4 keV predict a slighty higher 70/24 $\mu$m
ratio of 0.58.) We then derive a total dust 
mass of $5.4 \times 10^{-4}$ $M_\odot$ from the measured spatially-integrated 
MIPS fluxes. 

We obtain nearly the same mass when we use instead plasma 
conditions assumed by Douvion et al. (2001), $n_e = 6000$ cm$^{-3}$ and 
$kT_e = 0.0043$ keV, corresponding to a model spectrum shown by a solid line 
in their Figure 3, and based on fits to
IRAS and ISO observations. (Douvion et al. 2001 quote a smaller dust mass of 
$1-2 \times 10^{-4}$ $M_\odot$, appropriate for their simple model with hot 
silicate 
dust at temperature of 107.5 K, significantly hotter than dust in our 
models.) This 
agreement between two different mass determinations based on independent
data and very different assumed plasma conditions is encouraging; 
a future more detailed spatially-resolved joint study of {\it Spitzer} and 
{\it Chandra} data should provide us with a refined dust mass 
determination. The mass of $5.4 \times 10^{-4}$ $M_\odot$ refers to dust 
currently present in the shocked CSM. Most (78\%) of dust was destroyed in 
our fast shock model, implying an initial (preshock) dust mass of 0.0024 
$M_\odot$. Because dust destruction rates depend sensitively on the assumed
shock speed and its age, our current estimate of the preshock dust mass is
rather uncertain.

The total IR flux in the plane shock model discussed above is 
$1.4 \times 10^{-9}$ ergs s$^{-1}$ cm$^{-2}$, in good agreement with 
IRAS-based fluxes of $1.3 \times 10^{-9}$ ergs s$^{-1}$ cm$^{-2}$ and 
$1.6 \times 10^{-9}$ ergs s$^{-1}$ cm$^{-2}$ listed by Dwek (1987)
and Arendt (1989), respectively. The IR luminosity is
$2.8 \times 10^{36}$ ergs s$^{-1}$. Kepler is a low-luminosity object when 
compared with SNRs for which this type of analysis has been 
done with IRAS (e.g., Saken et al. 1992). More recently, Borkowski et 
al. (2006) derived total luminosities for two 
of the four SNRs from Type Ia SNe in their study, and B. Williams 
et al. (2006) derived luminosities for all four of their sample of 
SNRs from core-collapse SNe in the Large Magellanic Cloud. Of these six 
remnants, only SNR 0548-70.4 has a luminosity lower than Kepler, and 
even it is comparable at $2 \times 10^{36}$ ergs s$^{-1}$. This
SNR is, however, several times larger (and thus likely older) than
Kepler.  Indeed, most of the remnants in these two studies have 
luminosities much higher than Kepler. 

A specific object of interest for comparison is Tycho, a Type Ia SNR of 
comparable age located at a 2 kpc distance. The IR flux of Tycho is 
$5 \times 10^{-9}$ ergs cm$^{-2}$ s$^{-1}$ (Arendt 1989), so its IR
luminosity is comparable to Kepler. Dust in Tycho is much cooler than in
Kepler (Arendt 1989, Saken et al. 1992), consistent with a much lower
ISM density around Tycho. SNR 0509-67.5, one of the remnants studied by
Borkowski et al. (2006) in the LMC, appears very similar to Tycho.   
While observations of light echoes (Rest et al. 2005) have placed the 
age of this SNR at $\sim$ 400 yr, a high shock speed deduced from 
optical and UV observations of Balmer-dominated shocks (Ghavamian et al.
2007) implies a low ambient ISM density. Its MIPS 24 $\mu$m flux is 16.7
mJy, about 2--4 times lower than what Tycho or Kepler would have at a
distance of 50 kpc. An upper limit to its dust mass is 0.0011 $M_\odot$,
and SNR 0509-67.5 is likely less luminous in the IR than Tycho or Kepler. 

Determination of the dust/gas mass ratio requires knowledge of the shocked
CSM gas mass. Using an X-ray emission measure of 10 $M_\odot$ cm$^{-3}$ 
(Appendix A) and $n_e = 13$ cm$^{-3}$, we 
derive a total shocked CSM mass in Kepler of 0.77 $M_\odot$. This is in 
good agreement with a shocked CSM mass of 0.95 $M_\odot$ derived from 
{\it ASCA} observations (Kinugasa \& Tsunemi 1999), after scaling 
their results to the 4 kpc distance used here and using an electron density of 
13 cm$^{-3}$. The hydrodynamical model of Borkowski et al. (1992, 1994), 
based on the massive core-collapse runaway progenitor scenario of 
Bandiera (1987), requires $\sim 1 ~ M_\odot$,
also in reasonable agreement with the present mass estimate. Recent 
hydrodynamical models of Vel\'{a}zquez et al. (2006), based on a Type Ia 
progenitor scenario, require several $M_\odot$ of shocked CSM.  But 
Vel\'{a}zquez et al. (2006) used
collisional equilibrium ionization X-ray spectral models, which
underestimate X-ray emission by an order of magnitude or more for young SNRs
(Hamilton et al. 1983; see also Appendix A), resulting in an excessive 
estimate of the shocked mass by a comparable factor.

Combined with the dust mass from above of 0.0024 $M_\odot$, we arrive at a
dust-to-gas mass ratio for the CSM surrounding Kepler of 0.003. This is 
lower than the generally accepted figure for the Galaxy (e.g., 
Weingartner \& Draine 2001) by a factor of several.  Since the dust 
in Kepler appears to have originated in some kind of a stellar outflow,  
the lower dust content may be related to this and not a characteristic
of the general ISM in the region. Kepler's 
dust/gas ratio is higher by a factor of several than found for SNRs 
in the Magellanic Clouds (Borkowski et al.~2006; B.~Williams et al.~2006; 
Bouchet et al. 2006), most likely because of the high (near or above solar) 
metallicity of its SN progenitor compared with the $\sim$0.4 
solar LMC abundance.

\subsection{North-South Density Gradient}

A number of indicators imply a potential difference in preshock density
from north to south across the Kepler region.
The difference in the surface brightness of the north rim compared to 
the south rim is one indicator, and the effect may be visible directly 
from the background in Fig. 1 and the appearance of the 160 
$\mu$m image in Fig. 2e. The south rim is
only faintly visible at 24 $\mu$m, while the north rim is extremely
bright. We believe the 24 $\mu$m is dominated by dust emission, 
although some contribution from emission lines cannot be ruled out 
without spectra.  Under this assumption, we took a small region of the southern 
rim and an equal-sized region of the northern rim that is
dominated by nonradiative shocks and measured the flux from both
regions at 24 $\mu$m. The regions were arcs following the shape of the
rim, approximately 23\arcsec\ in thickness and 75\arcsec\ in length. 
The south region follows the faint southern rim at a declination of 
approximately -21:31:04 (J2000).  We found a north/south ratio of 24 $\mu$m fluxes 
of $\sim$30. A possible explanation for this contrast is that the 
shocks are encountering regions of different density. 

We have made several estimates of the density contrast necessary to 
produce the observed intensity ratio.  First, we assume that the 
dust-to-gas mass ratio in the two regions is the same (although we make 
no {\it a priori} assumptions 
about what that ratio must be). We do not assume equal amounts of 
swept-up material in the two regions, but rather simply vary 
the gas density (and dust mass accordingly) in the models and assume 
the same shock speed (and thus the same temperature) for both regions. 
This method reproduces the observed flux difference with a modest 
density contrast of only a factor of $\sim$4.5. It should be noted 
that these are all post-shock densities.

As an alternative approach to this problem, one can relax the
requirement that the shock speeds be equivalent in the two regions,
since a variation in density should cause variations in shock speed.
We repeated the calculations for several
different shock speeds in the south region, while keeping the north
region constant at 1660 km s$^{-1}$, and a density of $n_{e} = 19.5$ 
cm$^{-3}$.  We calculated electron and proton temperatures for the 
south region with these different shock speeds assuming no equilibration
at the shock front, and used these different temperatures to predict 
dust emission at
24 $\mu$m.  In these models, the density contrast required is only 
weakly dependent on shock speed.  While grains
are heated to hotter temperatures in faster shocks, the amount of
destruction of grains also increases.  Overall, we find that for a
reasonable range of shock speeds up to 3000 km s$^{-1}$, a post-shock 
density contrast between 4 and 7 times lower in the south (and accordingly
4--7 times less dust present) can reproduce the observed ratio. The 
density contrast increases with faster shock speeds.

As a third approach, we consider the remnant to be
in pressure equilibrium.  Keeping the density and shock speed in 
the northern region constant with the values mentioned above, we 
varied the shock speed in the south and derived density from the
pressure
equilibrium expression $n_{N}V_{N}^{2} \equiv n_{S}V_{S}^{2}$,
where subscripts N and S refer to the north and south regions,
respectively.  Under this constraint, we determined the shock speed
required to reproduce the observed 24 $\mu$m flux ratio, again assuming
a constant dust-to-gas mass ratio. We find that a shock with speed 
$\sim$5000 km s$^{-1}$, and thus a density of $\sim$ 2.1 cm$^{-3}$ 
can account 
for the difference in fluxes coming from the two regions. This is a 
density contrast of 9.1.  Because 5000 km s$^{-1}$ is much higher 
than inferred from other indicators, this contrast in density is 
considered an upper limit. The predicted 70/24 $\mu$m flux ratio 
in the south region is $\sim$ 1.2, which is significantly higher 
than in the brighter regions of the remnant, but consistent with the
idea that the dust in that region is cooler.
It is also consistent in a general sense with the higher overall 
average 70/24 $\mu$m ratio of 0.52.  Since the these models are 
only constrained by the fluxes
observed at 24 $\mu$m in the south, these results should be
considered tentative.  Detailed mid-infrared spectroscopy will be
required to investigate this issue further.

\subsection{Whither the Cold Dust Component?}

Morgan et al. (2003; hereafter M03) have recently reported excess
emission in Kepler using SCUBA 450 and 850 $\mu$m observations. 
They model this as a cold astronomical silicate dust component from 
ejecta ($T_d$ = 17 K), inferring as much as a solar mass of material
in this component.  Dwek (2004a) proposes a differing interpretation,
arguing instead for a component at  $T_d$ = 8 K and a much lower mass of 
dust in the form of Fe needles.  He supports the general 
possibility of a cold dust component related to the ejecta, pointing to
a similar component identified in Cas A (Dwek 2004b), but the amount of 
mass involved would be much lower than inferred by M03. 
Douvion et al. (1991) would not have been able to detect this cold 
component.

Our MIPS 160 $\mu$m image shown in Figure 3 is directly relevant to 
this discussion since this wavelength is close to the peak of the 
cold component curve predicted by M03 (see their Figure 2). 
No emission related to the SNR shell
or interior is detected at 160 $\mu$m, although the background 
is patchy and there appears to be a general gradient of intensity 
from NE toward the SW in the image.

To derive an upper limit at 160 $\mu$m, we extracted emission from
the region corresponding to the SNR location (judging from an aligned
overlay of the 24$\mu$m image) and from nearby background regions, as 
described above for the 24 and 70 $\mu$m images.  Background levels 
to the north and east of the SNR are at a higher level than the average 
from the object location, 
while the background level in the south is slightly below the object 
region average.  The observed field of view is very close to the western
edge of the SNR, but the background in that region appears to be comparable 
to that in the south.  To be conservative, we apply the southern 
background level to the entire object, which will over-estimate
any contribution from the SNR.  The resulting upper limit we derive is
0.8 Jy at 160 $\mu$m, which is approximately a factor of 10 below the
value predicted by M03 from the SCUBA-based model. Hence, our $\spitzer$
160 $\mu$m data do not confirm the M03 cold-dust picture. This fairly 
conservative upper limit does not rule out the idea of Fe needles
as discussed by Dwek (2004a), but the presence of any such component in
Kepler has been argued against on theoretical grounds (Gomez et al. 
2005).  Even if present, any such component in Kepler's SNR would
contain well below 0.1 solar mass, and would not require a massive 
precursor star as in the M03 interpretation.

\subsection{Synchrotron Emission}

X-ray synchrotron emission has been reported for Kepler: thin
filaments seen in the eastern ``ear" extension (Bamba et al.~2005),
broader emission from the SE quadrant (Cassam-Chena\"i et al.~2004),
and an extension of the integrated flux to hard X-rays (Allen,
Gotthelf, \& Petre 1999).  The X-ray fluxes demand a steepening of the
extrapolated radio spectrum; if that steepening occurs at shorter than
IR wavelengths, IR synchrotron emission should be present with a
morphology identical to the radio and at readily extrapolated
brightness levels.  While the average radio spectral index of Kepler
is --0.71 ($S_\nu \propto \nu^\alpha$) (DeLaney et al.~2002),
higher-frequency archival radio data imply a value of --0.59 (Reynolds
\& Ellison 1992).  This concave-up curvature, or hardening to shorter
wavelengths, can occur for electrons accelerated in a shock modified
by the pressure of accelerated ions (Ellison \& Reynolds 1991).  If we
extrapolate from images at cm wavelengths (e.g., DeLaney et al.~2002)
with this value, we predict surface brightnesses at 3.6 $\mu$m in the
range 0.6 -- 1.1 $\mu$Jy arcsec$^{-2}$ (or $(2.5 - 4.8) \times
10^{-2}$ MJy sr$^{-1}$).  Our 3.6 $\mu$m image shows only a faint hint
of any emission associated with Kepler.  In particular, there is no
apparent emission in the SE quadrant.  Typical brightness levels are
0.4 -- 0.5 MJy sr$^{-1}$, or 10 -- 20 times the extrapolated
synchrotron flux.  Even with very significant hardening of the
spectrum above 10 GHz, the 3.6 $\mu$m data are not constraining the
presence of synchrotron emission.  In particular, the regions
identified by Bamba et al. (2005) and Cassam-Chena\"i et al.~(2004), 
the SE quadrant and the eastern ``ear," are very faint in IR in 
general.  The western ``ear" may also harbor X-ray synchrotron 
emission, and it too is fainter at 24 $\mu$m relative to the rest 
of the shell emission. Far more sensitive IR observations would be 
necessary to detect synchrotron emission.

\section{Discussion}


Of the historical SNe, Kepler's remains enigmatic because a clear
determination of the SN type (and thus precursor star) has proven 
elusive.  Claims of evidence supporting both a massive precursor 
(core collapse SN type) or a white dwarf precursor (Type Ia SN) 
abound in the literature over the last decade or more.  In this 
section, we discuss this issue and highlight new insights that may
help resolve this dichotomy.

Firstly, since not all SN ejecta can form into dust, the claim by 
M03 of a solar mass or more of cold dust in Kepler
was an indication of even more ejecta mass and hence a massive 
precursor star. The negative detection at 160 $\mu$m here points
to an apparent problem with the SCUBA result, presumably due to the
complex and variable background observed in this region.  Hence,
a massive precursor is not required by the IR data.

Another indication pointing toward a massive precursor star has been the
overabundance of nitrogen in the optically-emitting filaments, which are
thought to represent dense knots of CSM being overrun by the blast wave.
We have run a small grid of shock models using the current version of
the shock code described by Raymond (1979) and Hartigan et al. (1987)
for comparison with the optical spectrum of knot `D3,' a bright,
radiative shock knot presented by Blair et al. (1991, their Table 2).  
We find that the main features of this 
spectrum can be matched with a model similar to model E100 of
Hartigan et al. (1987), but with the N abundance increased by 0.5 dex.
This is indeed a significant enhancement over solar abundance.

However, solar abundances are the wrong reference point.  At a distance 
of $\sim$4 kpc and galactic coordinate G4.5+6.8, Kepler's SNR is nearly half way 
to the galactic center.  Rudolph et al. (2006) provide a summary 
of galactic abundance gradients, and for nitrogen find  an increase
of $\sim$0.3 dex for the assumed distance of Kepler's SNR.  Additionally,
inspection of their Figure 4 shows that the scatter in observed points 
around the best fit line for the gradient readily encompasses the 
required value of 0.5 dex enhancement in nitrogen at Kepler's distance
from the galactic center.  While the 
nitrogen abundance around Kepler's SNR may be enhanced relative to
its local surroundings, it is not required, and certainly the 
magnitude of any enhancement is considerably less than has been 
recognized previously.

We note, however, that this in no way negates the fact that the 
presence of dense material surrounding Kepler's SN is surprising
for an object nearly 500 pc off the galactic plane. This material must
have its origin in either the precursor star, or the precursor system
(if a binary of some type was involved).  Canonical wisdom says that
core collapse precursors are massive stars, and massive stars shed 
material via stellar winds prior to exploding.  Type Ia SNe are
thought to arise from white dwarf stars that are pushed over the
Chandrasekhar limit, although the exact details of the precursor
system that gives arise to this are still widely debated (e.g. 
Livio \& Riess 2003, and references therein).
The absence of hydrogen lines in the spectra of Type Ia SNe is
taken as evidence for CO white dwarfs with little or no photospheric
hydrogen, and certainly little if any CSM.  

Improved statistics on extragalactic SNe, however, are finding 
exceptions to this general scenario.
A small but growing class of bona fide (confirmed with spectra)
Type Ia SNe has been found that show narrow hydrogen lines at 
late times, indicating the presence of a CSM component close to
these objects. SN 2002ic is a recent example of this phenomenon.
Some authors dub these objects Type Ia/IIn (e.g., a Type Ia with
narrow hydrogen lines; Kotak et al. 2004)
while others denote such objects as a new SN type, IIa (i.e., a 
Type Ia with hydrogen lines; Deng et al. 2004).  Whatever the designation,
these objects demonstrate that some SNIa explosions can occur in
regions with significant CSM, albeit at distances closer in to the SN
than inferred for Kepler.  Any such close in component near Kepler's 
SN, if present, would have long since been overrun by the blast wave.

Direct evidence for the presence of (presumably) CSM material 
around another Type Ia SN was provided by {\it Hubble}
Space Telescope images of light echoes around SN 1998bu (Garnavich 
et al. 2001; Patat 2005).  Among several light echoes
detected in this SN, there is an echo generated by scattering off dust
located closer that 10 pc to the SN.  At the same time,
SN 1998bu has one of the most stringent upper limits on the density of
the stellar wind (Panagia et al. 2006). The presence of a detached CSM
(or perhaps an ISM shell swept-up by winds of the SN progenitor) is
likely in this nearby SN. 
These and similar observations open up the possibility that
Kepler's SNR represents a local example of this phenomenon.

Finally, it should be noted that recent detailed X-ray observations
and modeling (Kinugasa \& Tsunemi 1999; Cassam-Chena\"i et al. 2004) are 
consistent with Si and Fe-rich ejecta, but show no evidence for other 
enhancements seen in core collapse SNRs such as Cas A (Hughes et al. 
2000; Hwang \& Laming 2003).  The temperature structure
inferred within the X-ray ejecta, with Fe K peaking interior to
Fe L is reminscent of Type Ia SNRs such as SN1006 and Tycho 
(Hwang \& Gotthelf 1997; DeCourchelle et al. 2001).  These early results
are strengthened almost to the point of certainty by a recent
deep {\it Chandra} X-ray exposure on Kepler that will allow a detailed
assessment of abundances within the ejecta of Kepler (Reynolds
et al. 2006). The preponderance of evidence from the X-rays now 
points toward a Type Ia origin for the precursor of Kepler's SNR.

If Kepler represents a SNIa explosion in a region
with significant CSM, it would be important from two directions:
If plausible models
are put forward that can explain a Type Ia with significant CSM, applying
them to Kepler's SNR may provide a stringent test because of its 
proximity and wealth of supporting observational data. On the other hand,
Kepler is only one object, and observations of additional extragalactic 
examples of SNIa's with CSM in various forms may provide important 
clues about the frequency and/or the progenitor population of such 
explosions (e.g. Mannucci 2005).


\section{Conclusions}

We have presented {\it Spitzer} imaging of Kepler's supernova
remnant at 3.6, 4.5, 5.8, 8, 24, 70, and 160 $\mu$m wavelengths
and compared with data from other wave bands.
Emission associated with the remnant is obvious at
all except 160 $\mu$m, but emission in the two shortest-wavelength IRAC
bands is visible only at the locations of bright optical 
radiative shocks in the WNW.  However, at 24 $\mu$m, the entire periphery of 
the remnant can be seen, along with emission seen, at least in projection,
toward the interior.

To summarize, we find:

\begin{itemize}

\item The 24 $\mu$m emission is well correlated with the outer blast
wave as delineated by soft X-ray emission and by nonradiative
(Balmer-dominated) shocks seen in the optical. This is clearly emission
from dust heated by collisions in the X-ray emitting material.  It is
not well correlated either with ejecta emission (shown by strong Fe
L-shell emission in X-rays) or with dense regions containing radiative
shocks.  In particular, we find no evidence
for dust newly formed within the ejecta material.

\item The emission at 8 $\mu$m largely resembles that of the
brightest regions at 24 $\mu$m,
although contamination by line emission in the 8 $\mu$m is apparent
in the radiative shock regions.  This similarity indicates
that even short-wavelength emission originates from the same grain
population as that at longer wavelengths.  However, current models are
not yet able to describe emission from the implied population of small
grains.

\item The emission at 70 $\mu$m is similar to that at 24 $\mu$m, but a
higher background makes it difficult to discern the fainter southern
half of the remnant.  Where both 70 $\mu$m and 24 $\mu$m emission can 
be seen, 70/24 $\mu$m flux ratios for discrete regions range from 0.37 to 
0.49.  Lower values, implying higher temperatures, are correlated 
with brighter regions. A total object average ratio of 0.52 implies
that the fainter regions tend toward higher values of this ratio.

\item The SED spectrum indicates that lines make at most a small
contribution between 55 and 95 $\mu$m.  The spectrum, along with the
absence of emission at 160 $\mu$m, rules out the presence of
large amounts of cold dust.  

\item Dust models using a power-law grain size distribution and
including grain heating and sputtering by X-ray emitting gas can
explain observed 70/24 $\mu$m flux ratios with sensible parameters.  
The models give gas densities of 10 -- 20 cm$^{-3}$.  The range of
observed ratios can be explained by ranges of temperature and density of
different regions in rough pressure equilibrium.

\item We find a total dust mass of about $5.4 \times 10^{-4}\
M_\odot$ after sputtering, and infer an original mass of about
$2.4 \times 10^{-3} \ M_\odot$.  With an estimate of shocked gas mass from X-ray
data, we infer an original dust/gas ratio of about $3 \times 10^{-3}$,
lower by a factor of several than normally assumed for the Galaxy, as
has been found for several other supernova remnants.

\item We find that a moderate density contrast in the range of
$\sim$4 -- 9 is required to explain the brightness variations
observed between the north and south rims of the remnant at 24 
$\mu$m, depending somewhat on the assumptions and models applied.

\item We suggest the preponderance of current evidence from 
optical, X-ray, and infrared data and modelling now points toward 
a Type Ia supernova, albeit in a region of significant surrounding 
CSM/ISM, especially for an object so far off the galactic plane.
A possible similarity to several extragalactic Type Ia supernovas
with narrow hydrogen lines at late times is pointed out.

\end{itemize}

Data from the {\it Spitzer Space Telescope} clearly demand more
sophisticated grain modeling.  The absence of large quantities of
newly formed dust challenges models hypothesizing such dust
formation in the SN ejecta.  However, the heated dust from the 
circumstellar medium heated by the blast wave can provide useful 
diagnostics of plasma conditions.

\acknowledgments

It is a pleasure to thank the $\spitzer$ operations team at 
JPL for their efforts in obtaining these data.  We also thank the
public relations staffs at the $\spitzer$, STScI, and $\chandra$ operations
centers for producing the color image in Figure 6, which was part of a 
photo release for the 400th anniversary of SN1604 in October 2004. 
This research has made use of SAOimage {\tt ds9}, developed by the
Smithsonian Astrophysical Observatory.  This work is supported
by JPL grant JPL-1264303 to the Johns Hopkins University.



\section{Appendix: X-ray Emission Measure of the Shocked CSM} 

We have used archival {\it XMM-Newton} data to estimate the 
emission measure of the shocked CSM around Kepler's SNR. While
Kepler's X-ray spectrum is dominated by ejecta
emitting strongly in lines of heavy elements such as Fe, Si, and S,
Ballet (2002) noted a good match between an RGS1 image around the O
Ly$\alpha$ line and the optical H$\alpha$ images. This suggests that
the low-energy X-ray emission dominated by N and O originates in the
shocked CSM. The Ly$\alpha$ lines of N and O, and the He$\alpha$ line
complex of O, are well separated from strong Fe L-shell lines in the
RGS spectra (see Figs. 1 and 2 in Ballet 2002). We used these
spectra to arrive at the following N and O line fluxes: $2.5 \times
10^{-13}$ ergs cm$^{-2}$ s$^{-1}$ for \ion{N}{7} $\lambda$24.779, $7.5
\times 10^{-13}$ ergs cm$^{-2}$ s$^{-1}$ for the \ion{O}{7} He$\alpha$
complex at $\sim 21.7$ \AA, and $2.8 \times 10^{-12}$ ergs cm$^{-2}$
s$^{-1}$ for \ion{O}{8} $\lambda$18.967.  The measured \ion{O}{8}
$\lambda$18.967 flux may include a non-negligible contribution from
\ion{O}{7} He$\beta$ $\lambda$18.627, as these two lines blend
together in the RGS spectra.

We have used a nonequilibrium-ionization (NEI) thermal plane-parallel shock
without any collisionless heating at the shock front to model N and O
line fluxes. This plane shock model is available in XSPEC as {\tt
vnpshock} model (Arnaud 1996; Borkowski et al. 2001). We 
assumed that N and O
lines are produced in fast (2000-2500 km s$^{-1}$; Sollerman et al. 2003) 
nonradiative, Balmer-dominated shocks with a
mean post-shock temperature of $\sim 5$ keV. The ISM extinction
$E(B-V)$ toward Kepler is equal to 0.90 (Blair et al. 1991), and 
with $R_V
= A_V/E(B-V) = 3.1$ and $N_H = 1.79 \times 10^{21} A_V$ cm$^{-2}$
(Predehl \& Schmitt 1995), $N_H$ is equal to $5.0 \times 10^{21}$
cm$^{-2}$. We assumed solar abundances for O 
(from Wilms et al. 2000).  We can reproduce \ion{O}{7} He$\alpha$ and
\ion{O}{8} $\lambda$18.967 (+ \ion{O}{7} He$\beta$ $\lambda$18.627)
fluxes with an emission measure $EM = n_e M_{g}$ equal to $10
M_\odot$ cm$^{-3}$ (at the assumed 4 kpc distance) and a shock
ionization age of $10^{11}$ cm$^{-3}$ s. By matching the measured
\ion{N}{7} $\lambda$24.779 flux, we arrive at an oversolar (1.6) N
abundance, confirming again the nitrogen overabundance in the CSM 
around Kepler.

The shock ionization age cannot be estimated reliably based on the
measured N and O line fluxes alone. These lines are produced close to
the shock front, and their strengths depend only weakly on the shock
age. Shocks with different ages can satisfactorily reproduce the
observed O and N line fluxes, with the emission measure $EM$ inversely
proportional to the shock age in the relevant shock age range from $5
\times 10^{10}$ cm$^{-3}$ s to $4 \times 10^{11}$ cm$^{-3}$ s.
Additional information is necessary to constrain the shock ionization
age, such as fluxes of Ne and Mg lines or the strength of the
continuum at high energies.  Because the X-ray spectrum of Kepler is
dominated by ejecta at higher energies, it is very difficult to
separate CSM emission from the ejecta emission in the
spatially-integrated {\it XMM-Newton} spectra. We note, however, that
shocks with ages of $2 \times 10^{11}$ cm$^{-3}$ s and longer produce
more emission than seen in Kepler.  In particular, the Mg lines and
the high energy continuum are too strong. For ionization ages as short
as $5 \times 10^{10}$ cm$^{-3}$ s, the O He$\alpha$/O Ly$\alpha$ line
ratio becomes excessive (0.36) in the model versus observations
(0.28).  Such short ionization ages may still be plausible if shock
ages are as short as $\sim 80$ yr and postshock electron densities are
$\sim 20$ cm$^{-3}$ as implied by infrared data.

We conclude that a reasonable estimate of the CSM emission measure,
equal to $10 M_\odot$ cm$^{-3}$, is provided by a plane shock with an
age of $10^{11}$ cm$^{-3}$ s. However, the $EM$ is known only 
within a factor of 2 because of the poorly known shock ionization
age and uncertain (perhaps spatially varying) absorption. (A 10\%\
range in absorption listed by Blair et al. (1991) results in 
35--40\%\ error in the derived EM.) 
It is also possible, even likely, that X-ray emission is produced
in a variety of shocks with different speeds driven into gas with
different densities. In this case, low energy X-ray emission would be
predominantly produced in slow shocks, while high energy emission
would originate in fast shocks. A single shock approximation used here
might then underestimate the true CSM emission measure. A future X-ray
study based on available high spatial-resolution {\it Chandra} data
will help in resolving such issues, and will result in better
estimates of the CSM emission measure.




\pagebreak
\centerline{\bf Figure Captions}
\bigskip
%
\noindent
\figcaption{ Full field of view MIPS 24 $\mu$m image of Kepler's SNR.
Scaling is set to a compromise level to show the overall structure
to best advantage. The 24 $\mu$m image is by far the deepest and most
detailed of the {\it Spitzer} images.}

\noindent
\figcaption{ A six-panel color figure concentrating on the IRAC images and 
their comparison to other wavelength bands. Panel a shows a three-color
IRAC image with 8 $\mu$m in red, 5.6 $\mu$m in green and 3.6 $\mu$m 
in blue.  Panel b is similar, but for 5.6 $\mu$m in red, 4.5 $\mu$m 
in green, and 3.6 $\mu$m in blue.  The orange color of the SNR filaments 
indicates emission in both 4.5 and 5.6 $\mu$m, but only from the 
brightest filaments seen at 8 $\mu$m.  Panel c is a difference image 
of 8 $\mu$m minus 5.6 $\mu$m, scaled to show the extent of faint 
emission at 8 $\mu$m.  Panel d shows the 4.5 $\mu$m minus 3.6 
$\mu$m difference image.  Panel e shows the 24 $\mu$m MIPS 
image from Fig. 1 to the same scale as the other images.  The
8 $\mu$m image closely tracks the brightest regions at 24 $\mu$m.
Panel f shows the soft band (0.3 - 0.6 keV) {\it Chandra} image from 
archival data, which again looks astonishingly like the 8 $\mu$m 
image in panel c.  All images are aligned and scaled exactly the same.}

\noindent
\figcaption{ A six-panel color figure concentrating on the MIPS images and 
their comparison to other wavelength bands. Panels a and b show the MIPS 24 
$\mu$m data with a hard stretch and a soft stretch, respectively, to 
show the full dynamic range of these data.  Panel c shows the MIPS 70 
$\mu$m data after running through the GeRT software to improve the 
appearance of the background.  Panel e shows the MIPS 160 $\mu$m data 
for the same field of view, although no SNR emission is actually seen 
at this wavelength. Panel d shows the star-subtracted H$\alpha$ 
image from Blair et al. (1991).  Panel f shows a three-color 
representation of the {\it Chandra} data for Kepler, with the red being 
0.3 - 0.6 keV (as in Fig. 2f), green being 0.75 - 1.2 keV, and blue being
1.64 - 2.02 keV.}

\noindent
\figcaption{ SED apertures selected for assessing the bright NW
radiative emission and the sky background, projected
on the 24 $\mu$m image. } 

\noindent
\figcaption{ Background subtracted SED 55 - 95 $\mu$m spectrum of the NW region
of Kepler's SNR, as indicated in Figure 4.}

\noindent
\figcaption{ A color view of Kepler showing data from $\spitzer$ 24 $\mu$m 
(red), $\hubble$ ACS H$\alpha$ (yellow), and $\chandra$ medium (blue) and 
soft (green) X-ray emission bands.  Despite the differing intrinsic 
resolutions of the various data sets, they have been carefully 
coaligned.  A $\hubble$ continuum-band image was used to subtract the stars 
from this image.  See text for details.}

\noindent
\figcaption{ This figure shows the object and background 
extraction regions selected for determining ratios between 24 and 70 
$\mu$m (panels a and b) and between 8 and 24 $\mu$m (panels d and e).
For reference, the
regions are also projected onto the optical H$\alpha$ image from Blair 
et al. (1991) in panels c and f. Note that the images shown in panel a and d
are the versions that have been convolved to the lower resolution image.
Also, panel b shows the original (non-GeRT-corrected) 70 $\mu$m image.
The labels are used in the text and Tables 2 and 3.}

\noindent
\figcaption{ A ratio image of the 70 $\mu$m and 24 $\mu$m data, where only
the regions with significant signal have been kept.  The color bar provides 
an indication of the measured ratio, ranging from 0.28 to 0.9. A simple 
contour from the 24 $\mu$m image is shown for comparison.  Note the lower values 
of the ratio in the regions of brightest emission, indicating they are 
somewhat warmer.}

\noindent 
\figcaption{ The 70/24 $\mu$m MIPS flux ratio as a function of electron density and 
pressure for plane shock dust models discussed in the text. 
The background
color scale indicates 70/24 $\mu$m ratio, as indicated by the color bar
at right. Dashed lines are lines of  constant temperature. 
The solid magenta diagonal line at lower right indicates where the modeled shocks
would become radiative, assuming solar abundance models and an age of 400 years.
Three {\it solid curves} are lines of constant 70/24 MIPS flux 
ratios, 0.30, 0.40, and 0.52 (from top to bottom), encompassing 
measured ratios listed in Table 2 and the spatially-integrated ratio. 
Position of a Balmer-dominated fast 
(1660 km s$^{-1}$) shock in the north is marked by a star.}

\clearpage
\begin{figure}
\plotone{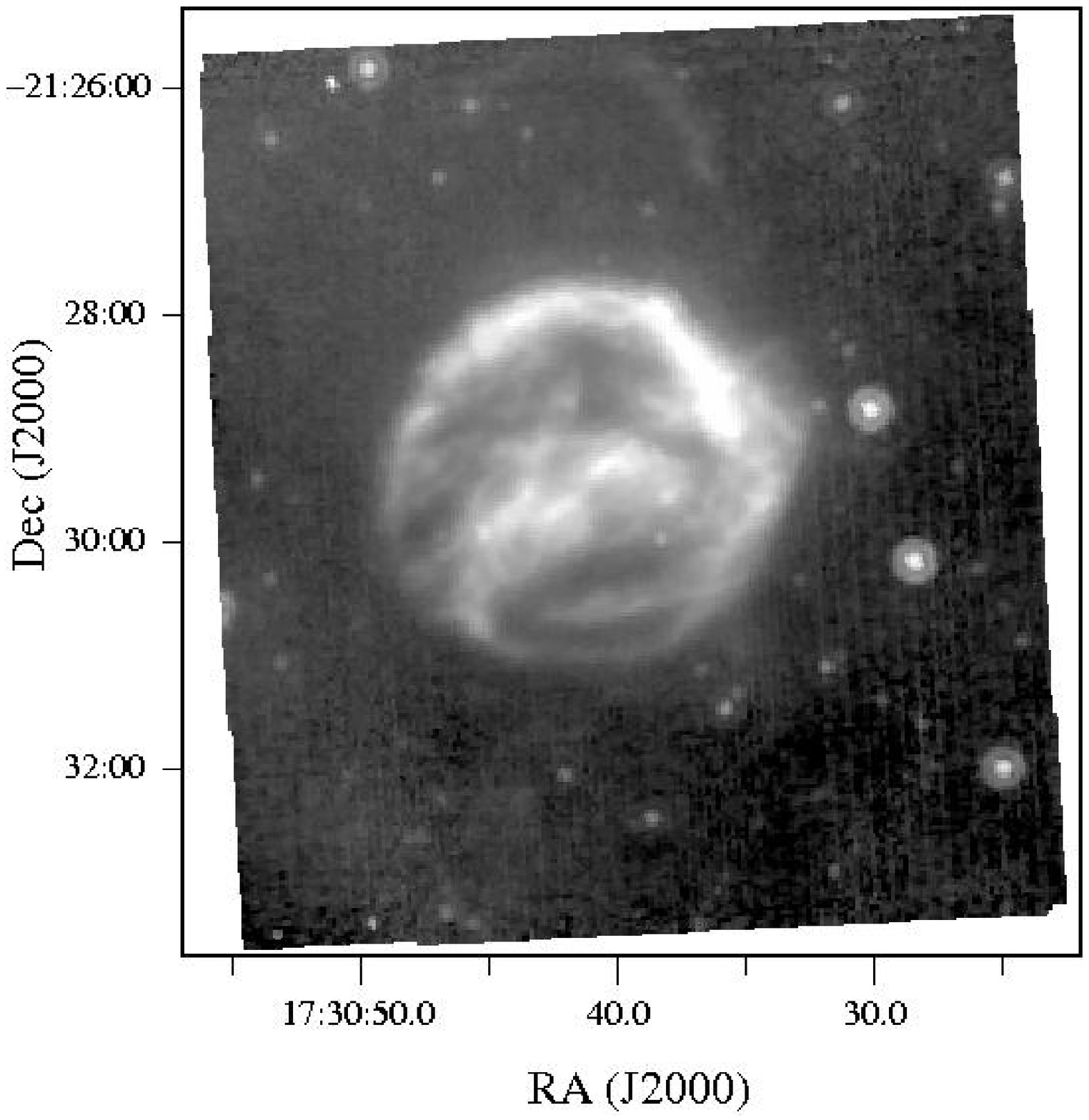}
\end{figure}

\clearpage
\begin{figure}
\epsscale{0.90}
\plotone{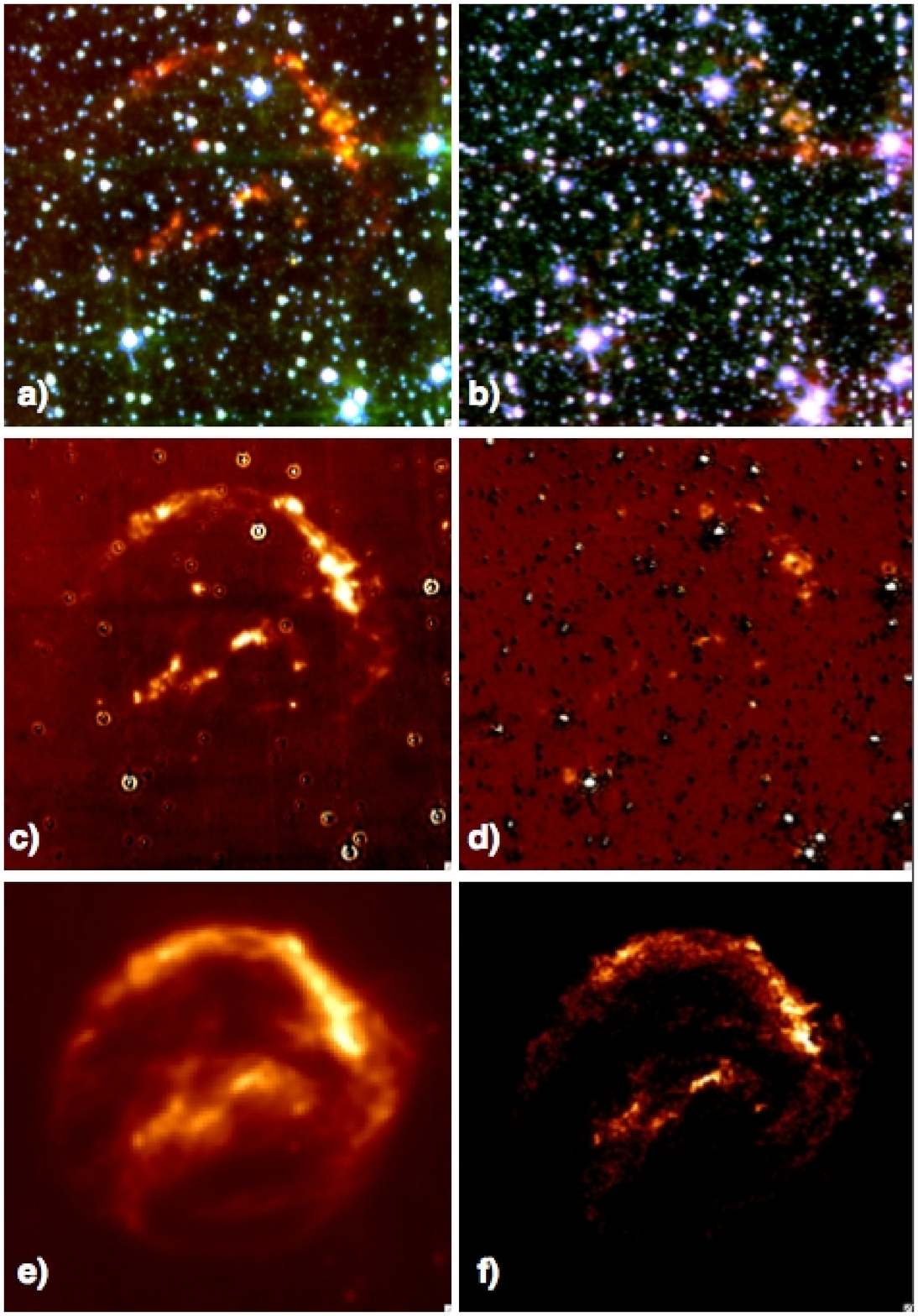}
\end{figure}

\clearpage
\begin{figure}
\epsscale{0.90}
\plotone{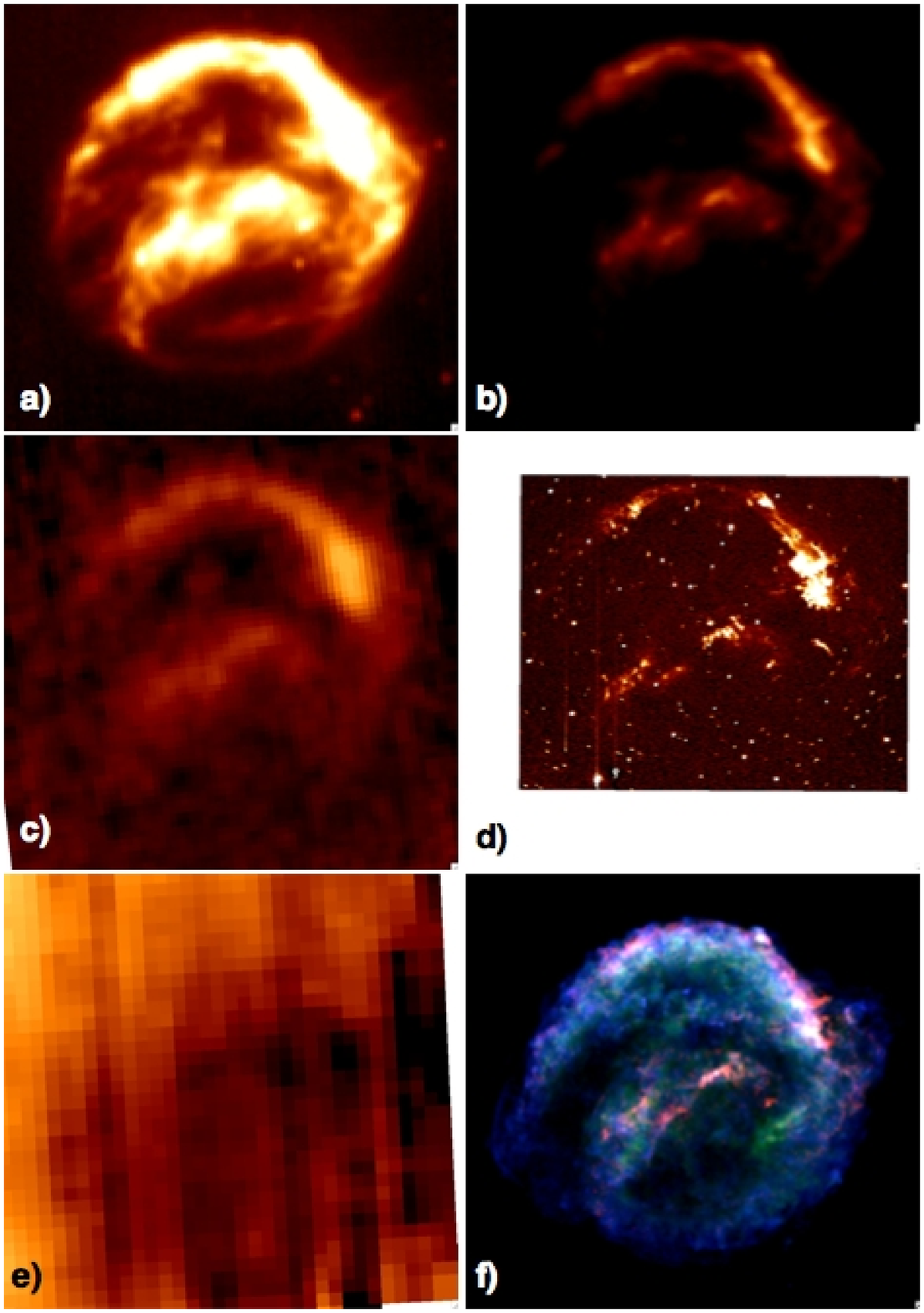}
\end{figure}

\clearpage
\begin{figure}
\plotone{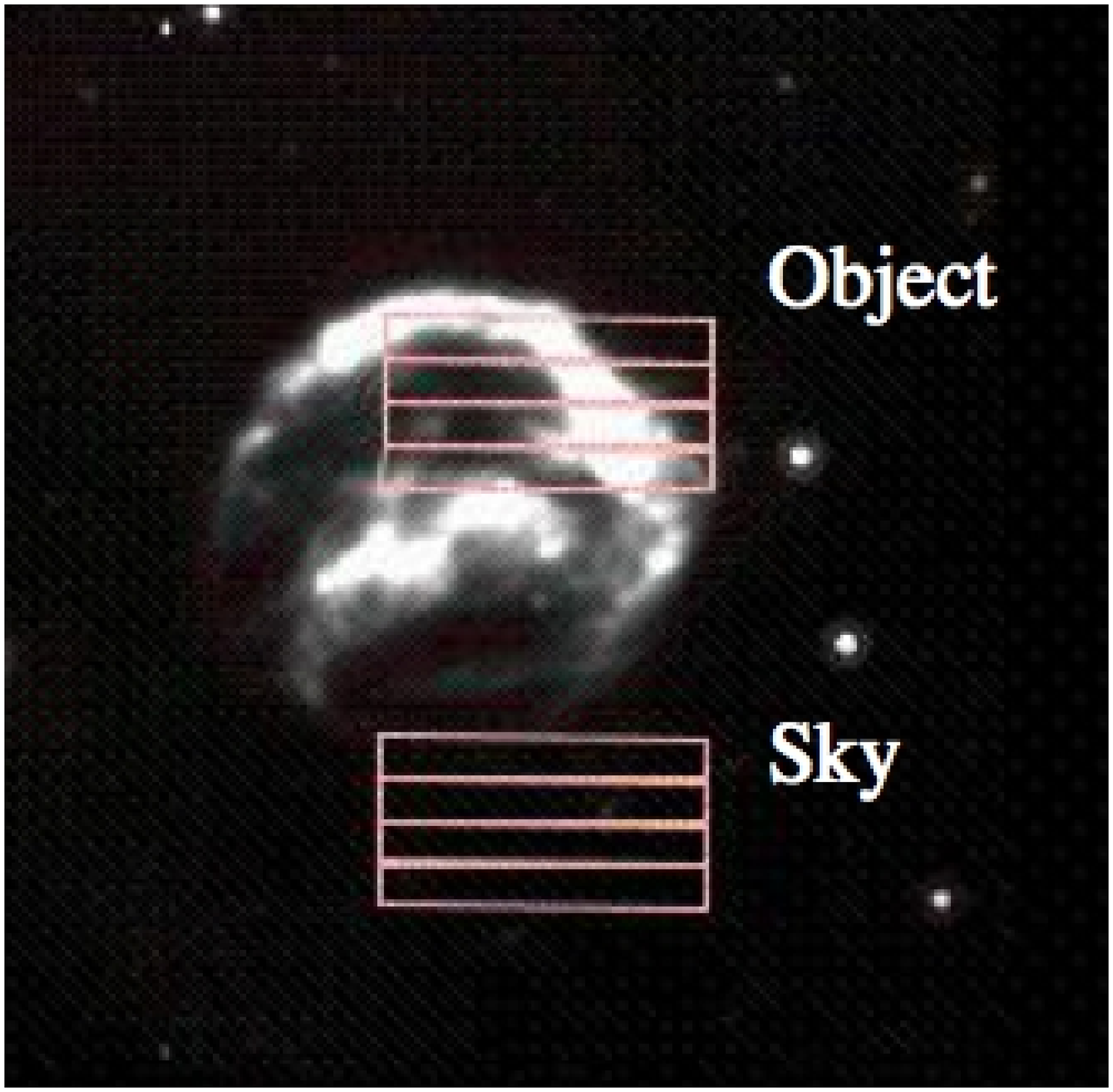}
\end{figure}

\clearpage
\begin{figure}
\plotone{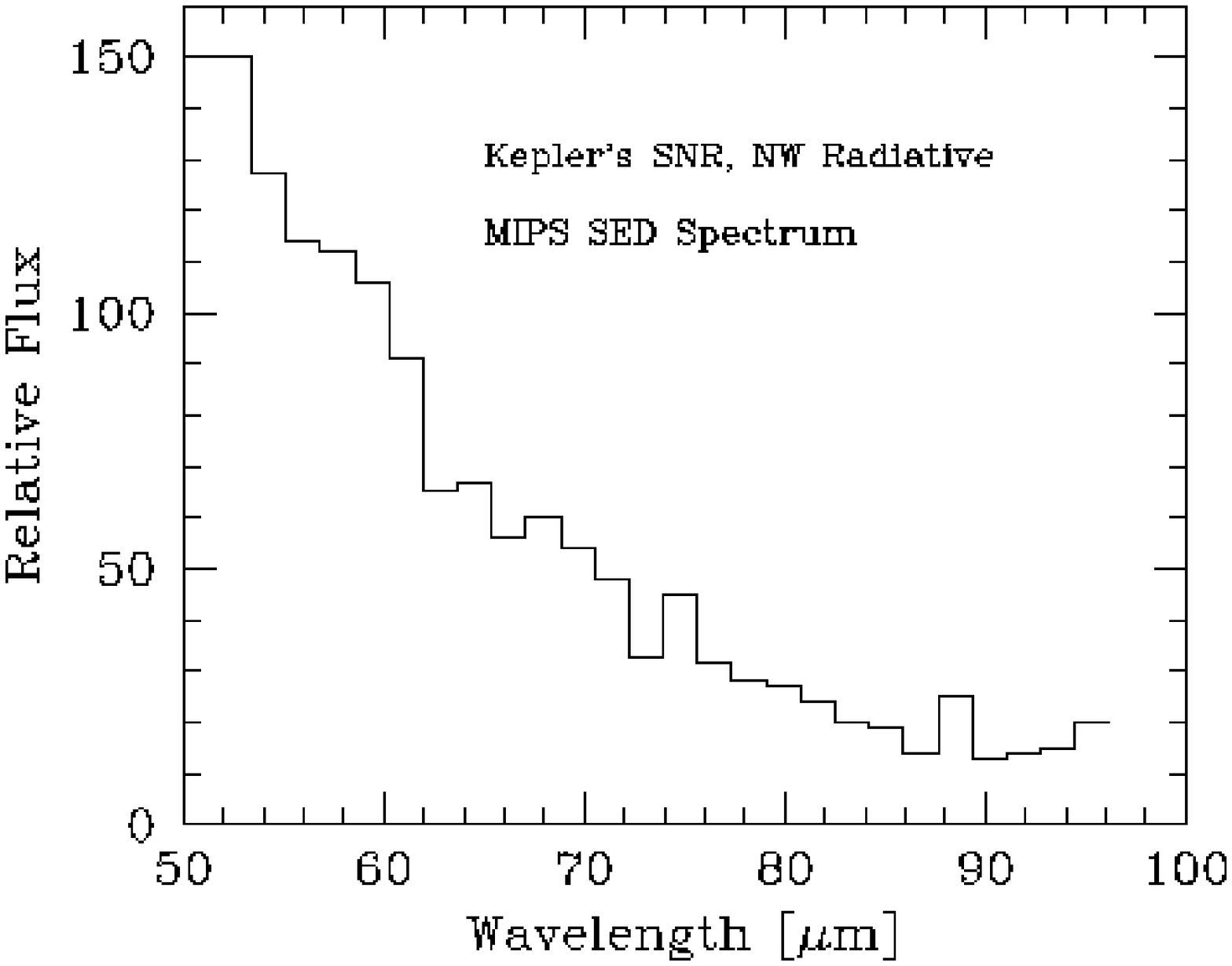}
\end{figure}

\clearpage
\begin{figure}
\plotone{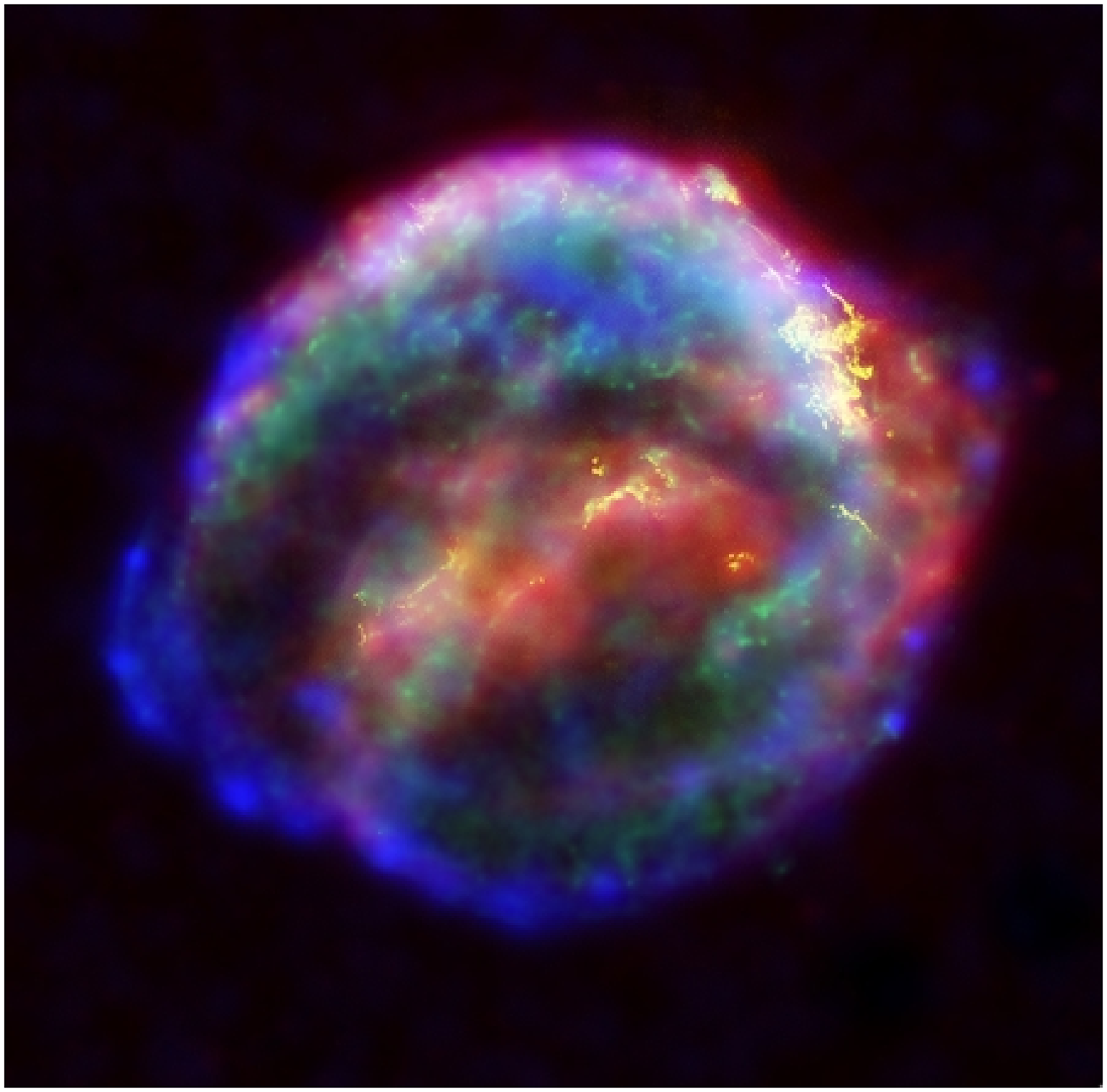}
\end{figure}

\clearpage
\begin{figure}
\plotone{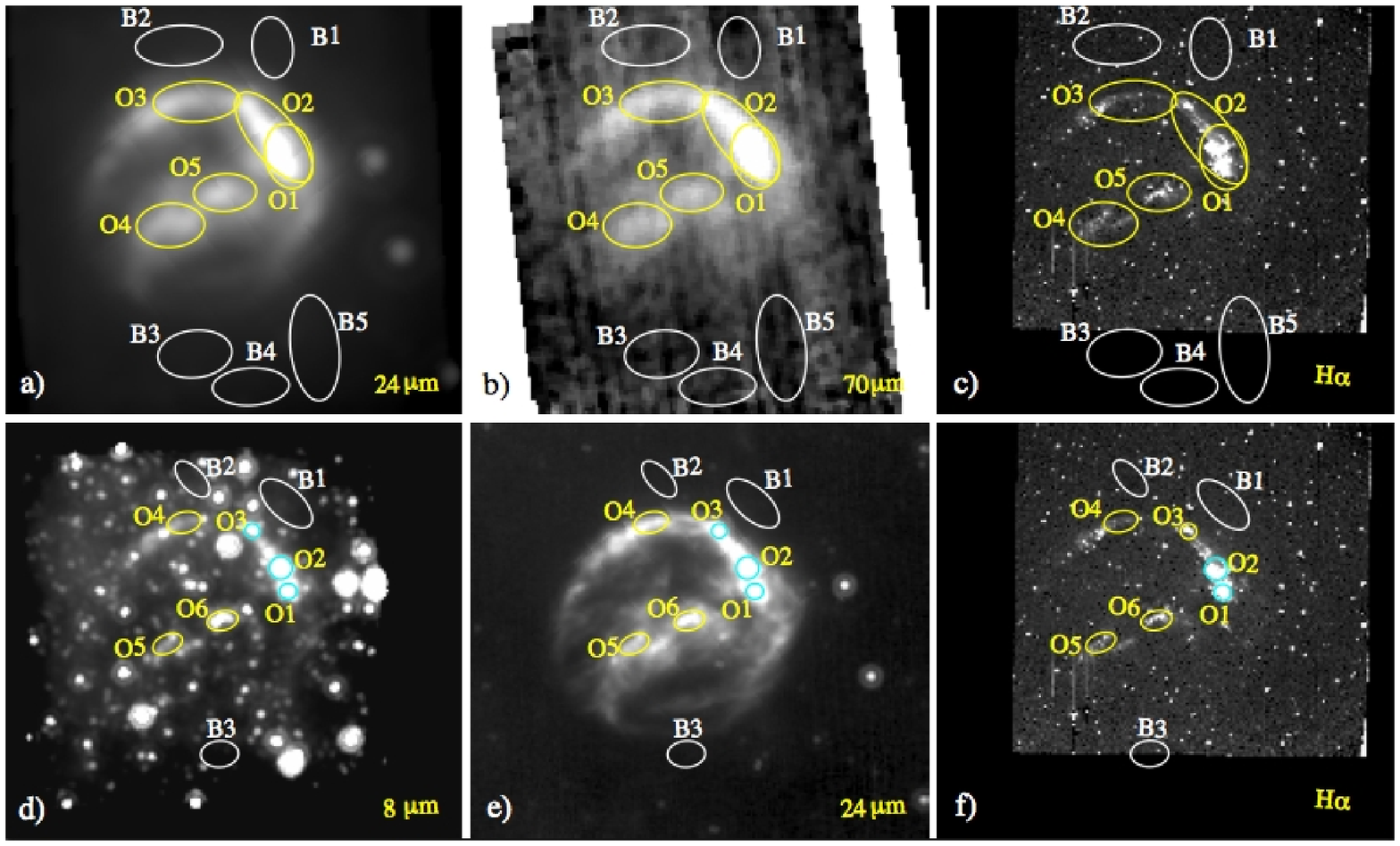}
\end{figure}

\clearpage
\begin{figure}
\plotone{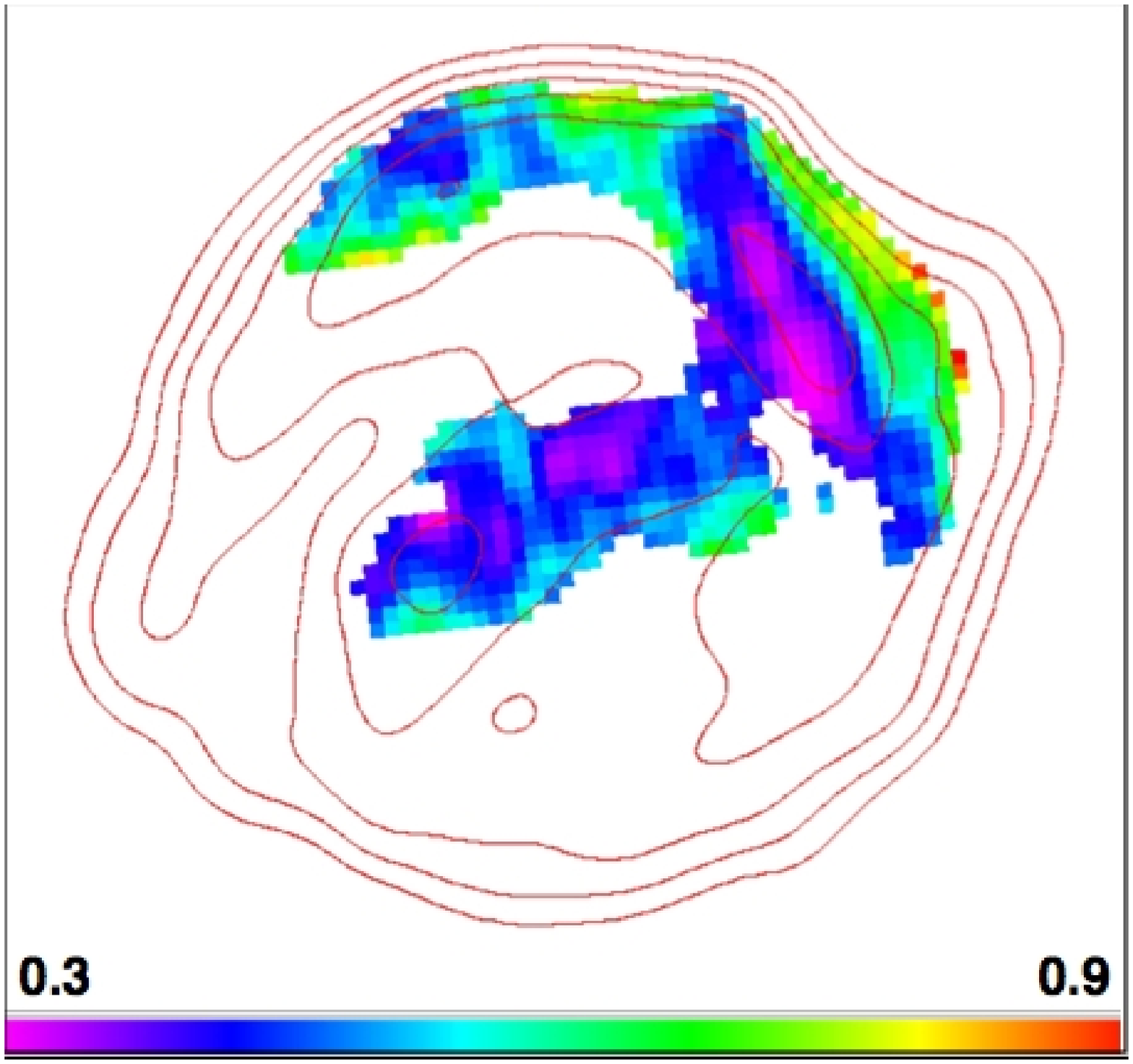}
\end{figure}

\clearpage
\begin{figure}
\plotone{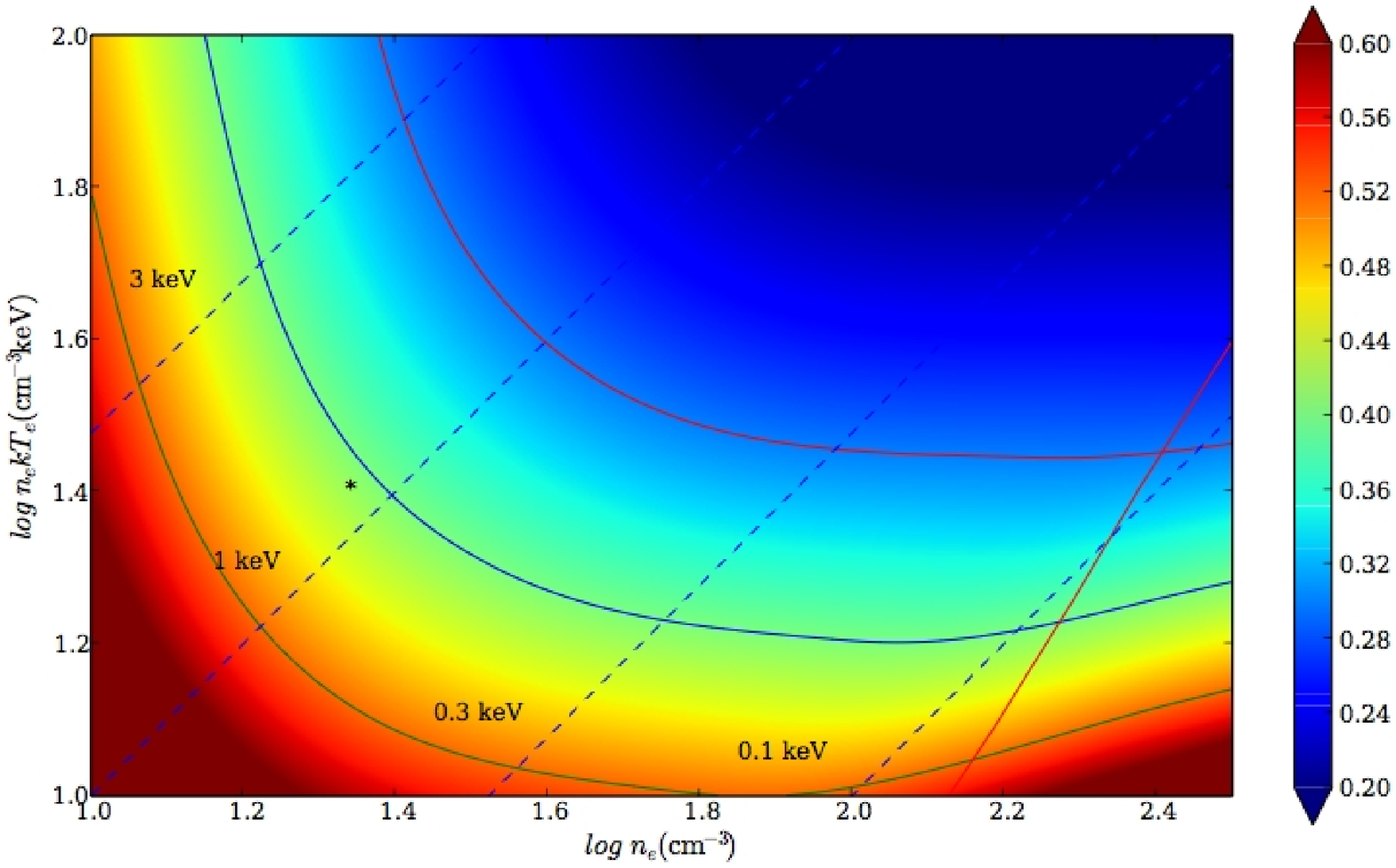}
\end{figure}



\begin{deluxetable}{lccccc}
\tablenum{1}
\tabletypesize{\footnotesize}
\tablewidth{5.5in}
\tablecaption{Aperture Parameters for Region Extractions\tablenotemark{a}}
\tablehead{\colhead{Region} &\colhead{$\alpha$ (2000) } &
\colhead{$\delta$ (2000) } &\colhead{Semi-Major } &\colhead{Semi-Minor  } &
\colhead{PA ($^{\circ}$) } \\
\colhead{ } & \colhead{ } & \colhead{ } &  
\colhead{Axis (\arcsec)} & \colhead{Axis (\arcsec)} &  \colhead{ }}
\startdata
{\bf 24 $\micron$ to 70 $\micron$: }  & &          &         &     &   \\
Object 1 & 17:30:35.76 & -21:28:44.5  &  21.4  & 30.0  & 192.9  \\
Object 2 & 17:30:36.73 & -21:28:27.2  &  20.8  & 50.2  & 218.9  \\
Object 3 & 17:30:41.81 & -21:27:54.6  &  39.8  & 18.4  & 182.9  \\
Object 4 & 17:30:43.51 & -21:29:48.0  &  31.2  & 20.2  & 182.9  \\
Object 5 & 17:30:39.93 & -21:29:17.9  &  28.8  & 17.2  & 182.9  \\
Background 1 & 17:30:36.82 & -21:27:05.2  &  18.5  & 27.9  & 186.3  \\
Background 2 & 17:30:42.96 & -21:27:03.2  &  39.8  & 18.4  & 182.9  \\
Background 3 & 17:30:41.91 & -21:31:45.7  &  34.0  & 21.6  & 186.3  \\
Background 4 & 17:30:38.24 & -21:32:15.7  &  35.1  & 17.2  & 182.9  \\
Background 5 & 17:30:34.05 & -21:31:40.2  &  22.3  & 48.7  & 186.3  \\
{\bf 8 $\micron$ to 24 $\micron$: }  & &          &         &     &   \\
Object 1 & 17:30:35.81 & -21:28:55.0  &  7.2  & 7.8  & 93.2  \\
Object 2 & 17:30:36.28 & -21:28:33.7  &  10.2  & 10.2  & 0  \\
Object 3 & 17:30:38.14 & -21:27:59.7  &  7.2  & 7.2  & 0  \\
Object 4 & 17:30:42.61 & -21:27:51.8  &  9.6  & 15.6  & 103.2  \\
Object 5 & 17:30:42.61 & -21:27:51.8  &  9.6  & 15.6  & 103.2  \\
Object 6 & 17:30:40.09 & -21:29:21.8  &  9.0  & 14.4  & 103.2  \\
Background 1 & 17:30:35.92 & -21:27:34.5  &  29.4  & 14.4  & 138.2  \\
Background 2 & 17:30:42.07 & -21:27:12.3  &  10.2  & 20.4  & 43.2  \\
Background 3 & 17:30:40.28 & -21:31:23.9  &  12.0  & 17.4  & 93.2 \\ 
\enddata
\tablenotetext{a}{Refer to Figure 7 for region identifiers projected
onto images.}
\end{deluxetable} 

\begin{table}
\begin{center}
{Table 2 -- MIPS 70/24 $\mu$m Regions Summary}
\begin{tabular}{lccc}
Region$^{a}$ & Net 70 $\mu$m$^{b}$ Flux & Net 24 $\mu$m Flux$^{b}$ & 70/24 Ratio \\
       & mJy & mJy &       \\
\hline
O1  & 778   &  2133  & 0.36 \\
O2  & 1171   &  3125 & 0.37 \\
O2 $-$ O1  & 394   &  992 & 0.40 \\
O3  & 412   &  1032 & 0.40 \\
O4  & 349   &  886  & 0.39 \\
O5  & 242   &  766  & 0.30 \\
\hline
\end{tabular}
\end{center}

Notes:  (a) Refer to Figure 7a-c for region definitions.
(b) Net fluxes have been background-subtracted as described in the text.
\end{table}


\begin{table}
\begin{center}
{Table 3 -- MIPS 8/24 $\mu$m Regions Summary}
\begin{tabular}{lccc}
Region$^{a}$ & Net 8 $\mu$m$^{b}$ Flux & Net 24 $\mu$m Flux$^{b}$ & 8/24 Ratio \\
       & mJy & mJy &       \\
\hline
O1  & 10.4   &  409   & 0.025 \\
O2  & 16.5   &  571   & 0.029 \\
O3  & 6.9    &  242   & 0.029 \\
O4  & 6.9    &  314   & 0.022 \\
O5  & 5.4    &  228   & 0.024 \\
O6  & 11.7   &  339   & 0.035 \\
\hline
\end{tabular}
\end{center}

Notes:  (a) Refer to Figure 7d-f for region definitions.
(b) Net fluxes have been background-subtracted as described in the text.
\end{table}



\begin{references}

\reference{}
Allen, G., Gotthelf, E., \& Petre, R.
1999, Proc.~26th ICRC, 3, 480

\reference{}
Arendt, R. G. 1989, ApJS, 70, 181

\reference{}
Arendt, R.~G., Dwek, E., \& Moseley, S.~H. 1999, ApJ, 521, 234

\reference{}
Arnaud, K.~A. 1996, in Astronomical Data Analysis and Systems V,
eds. G.Jacoby \& J.Barnes, ASP Conf. Series, v.101, 17

\reference{}
Baade, W. 1943, ApJ, 97, 119

\reference{}
Ballet, J. 2002, in High Resolution X-ray Spectroscopy with {\it
XMM-Newton} and {\it Chandra} (London)

\reference{}
Bamba, A., Yamazaki, R., Yoshida, T., Terasawa, T., \& Koyama, K.
2005, ApJ, 621, 793

\reference{}
Bandiera, R. 1987, ApJ, 319, 885

\reference{}
Bianchi, L., et al. 2005, ApJ, 619, 71

\reference{}
Blair, W. P. 2005, in ``1604-2004: Supernovae as Cosmological Lighthouses,''
ASP Conf. Ser. 342, ed. by M. Turatto, S. Benetti, L. Zampieri, \& W. Shea
(San Francisco: ASP) 416

\reference{}
Blair, W.~P., Long, K.~S., \& Vancura, O. 1991, ApJ, 366, 484

\reference{}
Borkowski, K.~J., Blondin, J.~M., \& Sarazin, C.~L. 1992, ApJ, 400, 222


\reference{}
Borkowski, K.~J., Harrington, J. P., Blair, W. P., \& Bregman, J. P. 1994,
ApJ, 435, 722

\reference{}
Borkowski, K. J., Lyerly, W. J., \& Reynolds, S. P. 
2001, ApJ, 548, 820


\reference{}
Borkowski, K.~J., et al. 2006, ApJ, 642, L141

\reference{}
Borkowski, K.~J., Sarazin, C.~L., \& Blondin, J.~M. 1994, ApJ, 429, 710

\reference{}
Bouchet, P., De Buizer, J. M., Suntzeff, N. B., Danziger, I. J.,
Hayward, T. L., Telesco, C. M., \& Packham, C. 2004, ApJ, 611, 394

\reference{}
Cassam-Chena\"i, G., et al. 2004, A\&A, 414, 545


\reference{}
Decourchelle, A., \& Ballet, J. 1994, A\&A, 287, 206

\reference{}
Decourchelle, A.,  et al. 2001, A\&A, 365, L218

\reference{}
DeLaney, T., et al. 2002, ApJ, 580, 914

\reference{}
Deng, J., Kawabata, K. S., Ohyama, Y., Nomoto, K., Mazall, P. A.,
Wang, L., Jeffery, D. J., Iye, M., Tomita, H., \& Yoshii, Y.  2004, 
ApJ, 605, L37


\reference{}
Doggett, J.~B., \& Branch, D. 1985, AJ, 90, 2303

\reference{}
Douvion, T., LeGage, P.~O., Cesarsky, C.~J., \& Dwek, E. 2001, A\&A, 373, 281

\reference{}
Draine, B.~T., \& Lee, H.~M. 1984, ApJ, 258, 89

\reference{}
Dunne, L. et al. 2003, Nature, 424, 285 

\reference{}
Dwek, E. 1987, ApJ, 322, 812

\reference{}
Dwek, E. 2004a, ApJ, 607, 848

\reference{}
Dwek, E. 2004b, ApJ, 611, 109

\reference{}
Dwek, E., \& Arendt, R. G. 1992, ARA\&A, 30, 11

\reference{}
Dwek, E., Foster, S.~M., \& Vancura, O. 1996, ApJ, 457, 244

\reference{}
Dwek, E., \& Smith, R. K. 1996, ApJ, 459, 686

\reference{}
Ellison, D.~C., \& Reynolds, S.~P. 1991, ApJ, 382, 242

\reference{}
Elmhamdi, A., et al. 2003, MNRAS, 338, 939

\reference{}
Fazio, G. G., et al. 2004, ApJS, 154, 10

\reference{}
Garnavich, P., et al.
2001, BAAS, 33, 1370

\reference{}
Gerardy, C.~L., \& Fesen, R.~A. 2001, AJ, 121, 2781

\reference{}
Ghavamian, P., Blair, W. P., Sankrit, R., Raymond, J. C., \&
Hughes, J. P. 2007, ApJ, submitted

\reference{}
Gomez, H. L., Dunne, L., Eales, S. A., Gomez, E. L., \&
Edmunds, M. G 2005, MNRAS, 361, 1021

\reference{} 
Guhathakurta, P. \& Draine, B. T. 1989, ApJ, 345, 230


\reference{}
Hamilton, A.~J.~S., Chevalier, R. A., \& Sarazin, C. L. 1983, ApJS, 51, 115

\reference{}
Hartigan, P., Raymond, J., \& Hartmann, L. 1987, ApJ, 316, 323

\reference{}
Hughes, J.~P. 1999, ApJ, 527, 298

\reference{}
Hughes, J. P., Rakowski, C. E., Burrows, D. N., \& Slane. P. O. 2000,  
ApJ, 528, L109

\reference{}
Hwang, U. \& Gotthelf, E. V. 1997, ApJ, 475, 665

\reference{}
Hwang, U. \& Laming, J. M. 2003, ApJ, 597, 362

\reference{}
Jurac, S., Johnson, R. E., \& Donn, B. 1998, ApJ, 503, 247

\reference{}
Kepler, J. 1606, De Stella Nova

\reference{}
Kinugasa, K., \& Tsunemi, H. 1999, PASJ, 51, 239

\reference{}
Kinugasa, K., \& Tsunemi, H. 2000, Adv. Sp. Res., 25, 559

\reference{}
Kotak, R., Meikle, W. P. S., Adamson, A., \& Leggett, S. K. 2004,
MNRAS, 354, L13

\reference{}
Krause, O., Birkman, S. M., Rieke, G. H., Lemke, D., Klass, U., 
Hines, D. C., \& Gordon, K. D. 2004, Nature, 432, 596


\reference{}
Livio, M., \& Riess, A. G. 2003, ApJ, 594, L93

\reference{}
Mannucci, F. 2005, in ``1604-2004: Supernovae as Cosmological Lighthouses,'' 
ASP Conf. Ser. 342, ed. by M. Turatto, S. Benetti, L. Zampieri, \&
W. Shea (San Francisco, ASP), 140

\reference{}
McCray, R. 1993, ARAA, 31, 175

\reference{}
Morgan, H.~L., Dunne, L., Eales, S.~A., Ivison, R.~J., \& Edmunds, M. G. 2003,
ApJ, 597, L33 


\reference{}
Panagia, N., Van Dyk, S. D., Weiler, K. W., Sramek, R. A., Stockdale, C.
J., \& Murata, K. P.
2006, ApJ, 646, 369

\reference{}
Patat, F. 2005, in ``1604-2004: Supernovae as Cosmological Lighthouses,'' 
ASP Conf. Ser. 342, ed. by M. Turatto, S. Benetti, L. Zampieri, \&
W. Shea (San Francisco, ASP), 229

\reference{}
Predehl, P., \& Schmitt, J. H. M. M. 1995, A\&A, 293, 889

\reference{}
Raymond, J. C. 1979, ApJS, 39, 1

\reference{}
Rest, A., et al. 2005, Nature, 438, 1132

\reference{}
Reynolds, S.~P., \& Ellison, D.~C. 1992, ApJ, 399, L75 

\reference{}
Reynolds, S.~P., et al. 2006, AAS 209, \#156.17

\reference{}
Reynoso, E.~M., \& Goss, W.~M. 1999, AJ, 118, 926


\reference{}
Rieke, G. H., et al. 2004, ApJS, 154, 25

\reference{}
Roellig, T.~L., \& Onaka, T. 2004, BAAS, 36, 1520

\reference{}
Rudolph, A. L., Fich, M., Bell, G. R., Norsen, T., Simpson, J. P., 
Haas, M. R., \& Erickson, E. F. 2006, ApJS, 162, 346

\reference{}
Saken, J. M., Fesen, R. A., \& Shull, J. M. 1992, ApJS, 81, 715

\reference{}
Sankrit, R. Blair, W.~P., DeLaney, T., Rudnick., L., Harrus, I. M., \& Ennis, 
J. A. 2005, Adv. Sp. Res., 35, 1027

\reference{}
Schaefer, B.~E. 1996, ApJ, 459, 438

\reference{}
Sollerman, J., Ghavamian, P., Lundqvist, P., \& Smith, R. C. 2003,
A\&A, 407, 249


\reference{}
Vel\'{a}zquez, P. F., Vigh, C. D., Reynoso, E. M., G\'{o}mez, D. O., 
\& Schneiter, E. M. 2006, ApJ, 649, 779

\reference{}
Weingartner, J. C., \& Draine, B. T. 2001, ApJ, 548, 296

\reference{}
Williams, B. J., et al. 2006, ApJ, 652, L33

\reference{}
Williams, R. M., Chu, Y-H., \& Gruendl, R. 2006, AJ, 132, 1877

\reference{}
Wilms, J., Allen, A., \& McCray, R. 2000, ApJ, 542, 914

\end{references}
\end{document}